\documentclass[sigconf]{acmart}
\AtBeginDocument{%
  }

\usepackage{listings}
\usepackage{tabularx} 
\usepackage{booktabs}
\usepackage{multirow}
\usepackage{xltabular}
\usepackage{array}
\usepackage[commandnameprefix=always]{changes}

\usepackage{longtable}
\usepackage{xcolor}
\usepackage{float}
\usepackage{tabularx} % For flexible-width columns
\usepackage[utf8]{inputenc}
\usepackage{textcomp}

\lstdefinestyle{promptstyle}{
  basicstyle=\ttfamily\small,
  columns=fullflexible,
  breaklines=true,
  frame=single,
  rulecolor=\color{black!20},
  showstringspaces=false,
  xleftmargin=0.5em,
  xrightmargin=0.5em,
  aboveskip=0.75em,
  belowskip=0.75em
}

\setcopyright{acmlicensed}
\copyrightyear{2018}
\acmYear{2018}
\acmDOI{XXXXXXX.XXXXXXX}
\acmConference[Conference acronym 'XX]{Make sure to enter the correct
  conference title from your rights confirmation email}{June 03--05,
  2018}{Woodstock, NY}
\acmISBN{978-1-4503-XXXX-X/2018/06}

\begin{document}

\title{AI and Collective Decisions: Strengthening Legitimacy and Losers’ Consent}

\author{Suyash Fulay}
\authornote{Both authors contributed equally to this research.}
% \author{Prerna Ravi}
% \email{prernar@mit.edu}
\affiliation{%
  \institution{Massachusetts Institute of Technology}
  \city{Cambridge}
  \state{Massachusetts}
  \country{USA}
}
\email{sfulay@mit.edu}

\author{Prerna Ravi}
\authornotemark[1]
\affiliation{%
  \institution{Massachusetts Institute of Technology}
  \city{Cambridge}
  \state{Massachusetts}
  \country{USA}
}
\email{prernar@mit.edu}

\author{Emily Kubin}
\affiliation{%
  \institution{Oxford University}
  \city{Oxford}
  \country{United Kingdom}}
\email{emily.kubin@psy.ox.ac.uk}

\author{Shrestha Mohanty}
\affiliation{%
  \institution{Massachusetts Institute of Technology}
  \city{Cambridge}
  \state{Massachusetts}
  \country{USA}
}
\email{shresmoh@mit.edu}

\author{Michiel Bakker}
\affiliation{%
  \institution{Massachusetts Institute of Technology}
  \city{Cambridge}
  \state{Massachusetts}
  \country{USA}
}
\email{bakker@mit.edu}

\author{Deb Roy}
\affiliation{%
  \institution{Massachusetts Institute of Technology}
  \city{Cambridge}
  \state{Massachusetts}
  \country{USA}
}
\email{dkroy@mit.edu}

\renewcommand{\shortauthors}{Fulay et al.}

\begin{abstract}
  
\end{abstract}




% \centering
%   \includegraphics[width=0.75\textwidth]{figures/teaser.png}
%   \caption{Our system visualizes a diversity of participant voices and enables \textit{hearing} policy-relevant lived experiences during decision-making.}
%   \label{fig:}

\received{20 February 2007}
\received[revised]{12 March 2009}
\received[accepted]{5 June 2009}

\begin{abstract}
% AI is often used to scale collective decision-making and identify consensus, but far less attention has been given to how it might foster trust, connectedness, and \added{\texit{losers’ consent}—the willingness of participants who do not get their preferred outcome to view the process and result as legitimate}. We present and evaluate a system designed to ground collective decisions in human voice and experience, thereby strengthening perceptions of legitimacy and social cohesion \added{specifically in cases where outcomes do not align with a participant's initial stance}. In a randomized experiment (n = 181) with Prolific workers, a subset of participants completed a 45-minute interview with an AI agent to share their beliefs and experiences on policy topics. These interviews generated an interactive visualization that displayed participants’ predicted policy support and allowed them to hear the voices and perspectives underlying those predictions. Compared to a no-visualization baseline, interacting with the system increased perceptions of legitimacy and cohesion. Our findings suggest that AI can not only improve the scale and quality of collective decision-making, but also foster trust, social connection, and \added{and losers' consent} throughout the process.

AI is increasingly used to scale collective decision-making, but far less attention has been paid to how such systems can support procedural legitimacy, particularly the conditions shaping losers’ consent—whether participants who do not get their preferred outcome still accept it as fair. We ask: (1) how can AI help ground collective decisions in participants’ different experiences and beliefs, and (2) whether exposure to these experiences can increase trust, understanding, and social cohesion even when people disagree with the outcome. We built a system that uses a semi-structured AI interviewer to elicit personal experiences on policy topics and an interactive visualization that displays predicted policy support alongside those voiced experiences. In a randomized experiment (n = 181), interacting with the visualization increased perceived legitimacy, trust, and understanding of others’ perspectives—even though all participants encountered decisions that went against their stated preferences. Our hope is that the design and evaluation of this tool spurs future researchers to focus on how AI can help not only achieve scale and efficiency in democratic processes, but also increase trust and connection between participants.

\end{abstract}
\maketitle

\section{Introduction}

Interest in using AI, particularly LLMs, to enhance collective decision-making has been growing rapidly. Here, collective decision-making involves groups forming preferences, exchanging information, and converging on shared outcomes whose legitimacy depends on acceptance even among those who disagree and “lose” \cite{davis2013group,list2011group,arrow1983collected}. Prior work shows LLMs can predict voting behavior \cite{gudino, yang, jarrett2025languageagentsdigitalrepresentatives, park2024generativeagentsimulations1000}, analyze large volumes of written opinions \cite{small2023opportunitiesrisksllmsscalable, fish2025generativesocialchoice}, and surface consensus at scale \cite{habermas, small2021polis}. 
These ideas are already deployed in practice (e.g., Pol.is) \cite{rodriguez2023taiwan, odonnell2025small}, but raise concerns that AI may replace—rather than augment—human participation, risking reduced human agency and civic disengagement \cite{garcia-marza2024algorithmic,roy2025beyond,barry}.

% One of the challenges in these processes is what political scientists call losers' consent: the willingness of participants whose preferred outcome does not prevail to nonetheless accept it as legitimate \cite{losers_consent}. Research suggests that participants’ perceptions of procedural fairness and their satisfaction with the deliberation process, rather than just the outcome itself, can influence their acceptance of decisions and future engagement \cite{stromer2009agreement, carman2010process}. Designing systems that support these process-level foundations may help increase the perceived legitimacy of collective decisions.

A key challenge in collective decision-making is what political scientists call “losers’ consent”: whether those whose preferred outcome doesn't prevail still accept it as legitimate \cite{losers_consent}. Accepting decisions and future engagement depends less on outcomes alone and more on perceived procedural fairness and deliberation quality, suggesting systems should prioritize these process-level factors \cite{stromer2009agreement,carman2010process}.
% Existing work in HCI has sought to improve the quality of deliberation through reflective nudges that encourage perspective-taking and understanding \cite{yeo2025enhancing, reflection_nudge, yeo2024helpmereflect}, as well as by mapping opinion spaces to help participants explore the pros and cons of different issues \cite{opinion_space, considerit}. Yet, these systems can sometimes ‘flatten’ participants’ contributions, reducing them to text-based opinions stripped of context about who they are and where their beliefs come from \cite{PolicyScape}.
% Additionally, while prior research has often prioritized efficient opinion aggregation and navigation, or finding consensus, less attention has been given to how the deliberation process \textit{itself} shapes participants’ perceptions of process legitimacy and social cohesion—factors central to achieving losers' consent.
Existing HCI work has focused on improving deliberation through reflective nudges that encourage perspective-taking \cite{yeo2025enhancing, reflection_nudge, yeo2024helpmereflect} and  mapping opinion spaces to explore issues' pros and cons \cite{opinion_space,considerit}, but this often flattens participants' contributions into decontextualized text stripped of their experiences and beliefs \cite{PolicyScape}. Also, prior work emphasizes efficient aggregation and consensus, overlooking how the deliberation process itself shapes legitimacy and social cohesion which are key to achieving losers’ consent.

% Accordingly, we focus our evaluation on these broader measures that capture the impact of participation beyond individual decision outcomes.

% \textcolor{red}{NEED TO ADD 2-3 SENTENCES ON CONNECTION TO CHI/HCI HERE
% sentence 1: what has HCI community been studying 
% sentence 2: what has it attempted to address the above problems 
% sentence 3: what gaps still remain? 
% }

Thus, rather than prioritizing consensus, we focus on process elements that help participants, especially dissenters, feel heard, respected, and understand others’ perspectives.
% Thus, rather than emphasizing consensus or optimization, we focus on the parts of the process that help participants, particularly those whose preferred outcome does not prevail, feel heard, respected, and better understand where others are coming from. 
Drawing on insights from deliberative democracy \cite{oxford_delib}, social psychology (particularly in bridging moral and political divides \cite{kubin, kessler2023hearing}), and HCI \cite{reflection_nudge, yeo2024helpmereflect, yeo2025enhancing, considerit, opinion_space}, we develop and evaluate a scalable collective decision-making system where decisions are supported by real, voice-based participant experiences to foster mutual understanding and trust in both process and outcomes.
% To achieve these goals, we first use a semi-structured AI interviewer to understand a participant's general background and experiences related to several topics. We then predict participants’ support for a given issue and provide an interactive tool where they can explore how people with different backgrounds, beliefs, and experiences relate to the topic across the full spectrum of predicted support.
We use a semi-structured AI interviewer to capture participants’ backgrounds on several topics, predict their support on given issues, and provide an interactive tool to explore how diverse beliefs and experiences relate to the topic across the full support spectrum.
Our main contributions are:
\begin{enumerate}
    \item A tool that supports losers’ consent by grounding outcomes in human voice and experience
    %A tool for scalable collective decisions grounded in human voice and experience
    \item An empirical study that evaluates this tool's impact on perceptions of process legitimacy, social cohesion, and learning, with a particular focus on supporting losers' consent
\end{enumerate}
% \textcolor{red}{NEED TO ADD RQs/Contributions HERE}

% 1. Tool (scalable)
% 2. Empirical study on two objectives
%     1. Process trust and legitimacy
%     2. Social cohesion

To evaluate our system, we ran a \(2 \times 2\) factorial randomized experiment ($n = 181$) with crowd-workers on Prolific on a series of work-related policy topics, where the two factors were participation in an AI interview and viewing a visualization. Topics included minimum wage increases, use of race and gender in hiring, and prioritizing domestic vs foreign workers.
% These topics were chosen because the general population has a diverse range of opinions on many of these issues and people may also have direct experiences with these topics \cite{Minkin2024_DEInegative, Dunn2021_MinWage}. Ninety of the participants took a semi-structured AI driven interview to collect their background information and experiences and beliefs related to the policies. These participants returned a week later to review the proposals and decisions, where 42 of these workers interacted with our tool visualization and 48 did not. Additionally, we had another set of 91 participants that did not take AI the interview, and split this group into those that did see our visualization of other participants' predicted stances and hear their input ($n = 44$) and those that did not ($n=47$). 
% Overall, we found that interacting with our visualization increased a range of pro-social measures. Participants were more likely to trust decisions, understand where they came from, and view them as legitimate. Additionally, they viewed other participants as more rational and were more likely to respect their positions. Interestingly, participants were not consistently more willing to interact or curious about others after using the tool; rather, exposure to different perspectives sometimes emphasized just how \textit{different} others were, suggesting the need for additional features that find commonalities and encourage interaction once the decision-making process is complete. 
Overall, the visualization increased pro-social outcomes: participants reported greater trust, understanding, and perceived legitimacy of decisions, and viewed others as more rational and worthy of respect. However, it did not consistently increase curiosity or willingness to engage with others, as exposure to differing views sometimes highlighted divisions—suggesting a need for features that surface common ground and encourage post-decision interaction.
% As AI is increasingly used in democratic practices, we show that, with appropriate design and agreement, these tools can not only improve decision quality but also strengthen understanding among participants. This holds true even when they disagree with one another or with the outcomes—an important but often understated goal of technology-assisted decision-making.
As AI becomes embedded in democratic practices, we show that well-designed systems can improve decision quality while strengthening mutual understanding among participants, even amid disagreement.

\section{Background}

\subsection{Increasing Trust in Collective Decisions}

\subsubsection{Lessons from Deliberative Democracy}
% How can we increase trust in collective decision-making? One inspiration comes from \textit{deliberative democracy}, where citizens exchange ideas and experiences before decisions are made \cite{oxford_delib}. Citizens’ assemblies, for example, have tackled contentious issues like abortion in Ireland and climate policy in the UK and Scotland \cite{fournier2011citizens, farrell2019systematizing, elstub2021scope, andrews2022emotional, cherry2021citizens, wells2021citizen}. Their success is often attributed to diverse representation, face-to-face interaction, and structured dialogue \cite{mutz2006hearing, karpowitz2014deliberation, gastil2005deliberative, setala2018mini}. Yet, these processes are limited in scale, typically involving only 100–200 participants. Our system cannot replicate in-person deliberation, but it seeks to preserve key benefits: amplifying human voice, fostering understanding of backgrounds, and exposing participants to different perspectives.

How can we increase trust in decision-making? Deliberative democracy offers approaches, where citizens exchange ideas and experiences before deciding \cite{oxford_delib}. Citizens’ assemblies show success through diverse representation, face-to-face interaction, and structured dialogue, but are limited in scale \cite{mutz2006hearing, karpowitz2014deliberation, gastil2005deliberative, setala2018mini}. While our system cannot replicate in-person deliberation, it aims to preserve its key benefits by amplifying human voice and exposing participants to diverse perspectives.

% We want to emphasize that many key characteristics of deliberative practices, such as back and forth interaction and facilitated dialogue, are not present in our system currently and hope that we can integrate some of these features in the future.

\subsubsection{Bridging Moral and Political Divides}
% Polarized conflicts often arise from deep moral divides, where people are unwilling to hear opponents’ perspectives and may even become more entrenched when exposed to them \cite{skitka2014social, frimer2017liberals, bail_cross_cutting}. Still, intervention research highlights strategies that can reduce polarization. Learning about opponents as individuals through their personal experiences has been shown to foster tolerance and reduce dehumanization \cite{kubin2023reducing, kessler2023hearing, kubin2025political}. Other effective approaches include helping people understand the reasoning behind opposing views \cite{puryear2024using}, emphasizing apolitical commonalities \cite{balietti}, and leveraging authentic voices rather than text \cite{schroeder2017humanizing}. Drawing on this literature, we designed our system to highlight a wide range of beliefs while preserving participants’ voices, backgrounds, and personal experiences.

Polarization often stems from moral divides that entrench opposing views \cite{skitka2014social,frimer2017liberals,bail_cross_cutting}. But intervention research shows exposing people to opponents’ personal experiences and reasoning behind their views \cite{puryear2024using}, emphasizing apolitical commonalities \cite{balietti}, and leveraging authentic voices over text \cite{schroeder2017humanizing} can foster tolerance and reduce dehumanization \cite{kubin2023reducing, kessler2023hearing, kubin2025political}. Accordingly, our system highlights a wide range of beliefs while preserving participants’ voices and backgrounds.

\subsubsection{Losers’ Consent}
% One challenge in collective decision-making is whether participants whose preferred outcome does not prevail still accept the decision as legitimate \cite{losers_consent}. Research suggests that such acceptance may depend more on the procedural fairness of the deliberation process than on the outcome itself \cite{zhang2015perceived}. Supporting these process-level aspects could help increase perceived legitimacy, particularly when participants disagree with the final decision. Our work focuses on fostering understanding and trust for those on the “losing” side of a decision.
A key challenge in collective decision-making is whether those who disagree still view decisions as legitimate \cite{losers_consent}. This depends more on perceived procedural fairness than outcomes. Our work thus focuses on fostering process-level understanding and trust for those on the “losing” side of a decision \cite{zhang2015perceived}.

\subsection{AI in Collective Decision-Making}

% LLM-enabled deliberation systems are shifting collective decision-making from preference aggregation toward facilitating reason-giving and negotiated understanding. Tools such as \textit{Pol.is} map opinion clusters and bridging statements at scale \cite{small2021polis,ovadya2023bridging}, and LLMs extend this by extracting and generating rationales and modeling preference structure across diverse populations \cite{bakker2022fine,small2023opportunities,burton2024how,habermas}. HCI scholarship cautions that “algorithm-in-the-loop’’ systems must preserve human agency, avoid automation bias, and remain accountable \cite{green2019principles}, leading recent work to position AI as a deliberative partner rather than a decision-maker \cite{ma2025humanai,fragiadakis2024evaluating}. Building on this shift, our system links model inferences to concrete, inspectable interview evidence, making predictions auditable and contestable.

LLM-enabled systems are shifting collective decision-making from preference aggregation to reason-giving and negotiated understanding. Tools like \textit{Pol.is} map opinion clusters and bridge statements at scale \cite{small2021polis,ovadya2023bridging}, while LLMs extract rationales and model preferences across diverse groups \cite{bakker2022fine,small2023opportunities,burton2024how,habermas}. Prior HCI work emphasizes preserving human agency and reducing automation bias \cite{green2019principles} by positioning AI as a deliberative partner rather than a decision-maker \cite{ma2025humanai,fragiadakis2024evaluating}. Our thus system grounds model predictions in inspectable interview evidence to support auditability.
% At the same time, sustaining pluralism remains crucial: premature consensus can silence disagreement and weaken legitimacy \cite{ovadya2023bridging}. We treat dissent as informationally valuable, surfacing diverse, high-relevance lived experiences rather than collapsing opinions into a single outcome. We evaluate whether such representations help participants feel heard and accept outcomes (i.e., “losers’ consent’’ \cite{losers_consent}), showing how LLMs can support not only aggregation but also trust, reflection, and pluralistic legitimacy in collective decision-making.
Simultaneously, sustaining pluralism is crucial, as premature consensus can silence disagreement and weaken legitimacy \cite{ovadya2023bridging}. We treat dissent as informationally valuable, use LLMs to surface diverse lived experiences instead of collapsing opinions into a single outcome, and evaluate whether this fosters losers’ consent by improving participants’ sense of being heard, reflection, and trust in outcomes \cite{losers_consent}. 

% \subsubsection{Interfaces that Surface Reasons and Support Reflection}
% Work in HCI emphasizes that decision-support systems should make not just outcomes but also underlying rationales visible, enabling people to evaluate fairness and legitimacy \cite{cheng2019explaining}. Interfaces that scaffold reflection—by prompting users to articulate reasons or compare perspectives—have been shown to measurably improve deliberative quality and mutual understanding \cite{yeo2024helpmereflect, considerit}. Our design builds directly on this work: predictions of policy support are paired with linked transcript and audio evidence, while guided exploration requires participants to engage in short reflective judgments. These mechanisms embed deliberative scaffolds into the workflow, aligning with HCI findings that transparency and reflection can mitigate automation bias and strengthen perceived process legitimacy \cite{arceneaux2017taming, reflection_nudge}. Additionally, our work builds on projects such as PolicyScape, which crowdsources diverse perspectives on policy issues and makes them accessible through an interactive exploration tool \cite{PolicyScape}, and ConsiderIt, which presents users with crowdsourced pro/con lists of policy proposals \cite{considerit}.  While previous work primarily focused on text-based opinions, our contribution extends this line of work by emphasizing the authentic voices, backgrounds, and experiences of participants, and by leveraging AI to capture these elements at scale.

\subsection{HCI and Collective Decision-making}

% The Human Computer Interaction (HCI) Community has long explored interfaces that make large-scale public input more navigable, comprehensible, and reflective—core requirements for legitimate collective decision-making. This work emphasizes that decision-support systems should make not just outcomes but also underlying rationales visible, enabling people to evaluate fairness and legitimacy \cite{cheng2019explaining}.  Early systems like \textit{Opinion Space} used deliberative polling to visualize a landscape of opinions and help users browse diverse comments; importantly, the spatial visualization increased users’ agreement with and respect for opposing views \cite{faridani2010opinion}. 
% This line of work established two design ideas we built upon: (1) mapping a spectrum of stances to make diversity legible, and (2) scaffolding exploration across that spectrum to broaden perspective-taking \cite{faridani2010opinion}. Our 2D visualization (see \autoref{fig:steps}) similarly positions participants across predicted support and explicitly curates profiles across the spectrum to encourage diverse exposure.

HCI has long explored interfaces that make large-scale public input navigable, interpretable, and reflective—key to legitimate decision-making—by surfacing underlying rationales alongside outcomes to help evaluate fairness \cite{cheng2019explaining}. Systems like \textit{Opinion Space} showed that mapping opinion spectra for exploration can increase agreement with and respect for opposing views \cite{faridani2010opinion}.
Multi-agent approaches have also been introduced with LLMs to model pluralistic deliberation \cite{ashkinaze2025plurals}.
Our 2D visualization (see \autoref{fig:steps}) similarly maps participants across predicted support and curates diverse profiles to encourage perspective-taking.

A parallel thread emphasizes \textit{structured reflection} to enhance deliberation quality \cite{yeo2025enhancing, yeo2024helpmereflect}. Prompting users to articulate pros/cons and engage with points raised by others can support more thoughtful reasoning \cite{considerit}. Beyond mere information access, reflection nudges that ask participants to clarify their views, adopt an opponent’s perspective, and infer others’ reasons increase attitude certainty and willingness to express opinions \cite{reflection_nudge}. 
Chatbot-mediated co-design shows that summarizing others’ inputs and prompting perspective-taking can increase willingness to commit to group decisions—even amid disagreement \cite{shin2022chatbots}. 
These increase deliberativeness \cite{yeo2024helpmereflect} and support higher-quality discourse in inflammatory online spaces \cite{yeo2025enhancing}. 
% Our system integrates this tradition by (a) using a semi-structured AI interview to elicit richer, experience-grounded positions before exposure to others’ opinions, and (b) embedding reflection prompts and Likert scales tied to each profile to focus users on understanding and developing empathy towards others’ views rather than rebuttal.
Building on this, our system elicits views and experiences via an AI interviewer and embeds reflection prompts to encourage understanding and empathy over rebuttal. 

% Another key insight is that \emph{who} is speaking matters. Prior work shows that platforms often surface opinions without contextualizing stakeholder identities, limiting understanding of lived consequences and intra-group variation \cite{PolicyScape}. Listening-centered designs such as \textit{Reflect} demonstrate that summarizing others’ viewpoints can increase communication satisfaction and willingness to participate \cite{kriplean2012you}. Our interface centers authentic voice by pairing profiles with lived experiences and transcript-linked audio, preserving tone while protecting anonymity. We also provide a self-view of users’ outward-facing profiles to enhance transparency and feeling heard \cite{kriplean2012you}, and, in line with Overney et al., support asynchronous participation, stance spectra, and prompts that surface both dominant and minority perspectives \cite{overney2025boundarease}.

Another factor is that platforms often surface opinions without contextualizing stakeholder identities, limiting understanding of lived consequences and intra-group variations \cite{PolicyScape}. Listening-centered designs such as \textit{Reflect} that summarize others’ viewpoints increase communication satisfaction and engagement \cite{kriplean2012you}. Our interface centers authentic voice by pairing profiles with lived experiences via transcript audio, adds a self-view to enhance transparency and feeling heard \cite{kriplean2012you}, and supports asynchronous participation while surfacing both dominant and minority perspectives \cite{overney2025boundarease}.

% Finally, HCI has begun to examine AI’s role in consensus-building. . More broadly, recent work explores LLMs as facilitators of reflective thinking rather than replacements for human voice \cite{yeo2024helpmereflect, yeo2025enhancing}, and multi-agent approaches model pluralistic deliberation \cite{ashkinaze2025plurals}. Our contribution advances this trajectory by pairing an AI interviewer (to scale experience-rich inputs) with an exploration interface that foregrounds human stories, provides reflection scaffolds, and closes the loop with transparent outcomes. 
% In evaluation, this design increased perceived legitimacy, trust, and respect for others’ rationality—outcomes prioritized but under-measured in prior HCI work focused on opinion aggregation or consensus alone \cite{faridani2010opinion, considerit, PolicyScape}.

% COMMENT THIS OUT WHEN DONE
% \input{HCI_notes}

\section{System Design and Implementation}
\subsection{Objectives}
The design of our collective decision-making system was guided by four primary objectives: ensuring \textit{process legitimacy}, fostering \textit{social cohesion}, supporting both individual and collective \textit{learning}, and enabling a broad \textit{scale of participation}. The first three were evaluated directly via surveys (\autoref{sec:measured_vars}), while \textit{scale of participation} informed our design of how users interacted with the system.

\subsubsection{Process Legitimacy}
A core aspect of sustainable democracy is viewing decision-making processes as fair and trustworthy \cite{dahl1989democracy, easton1965systems, why_trust_matters, levi2000political}. People accept decisions more readily when processes seem legitimate—described as "procedural justice" in the legal domain \cite{why_people_obey_law,lind1988social, thibaut1975procedural}—even when decisions go against their preferences \cite{losers_consent}.

\subsubsection{Social Cohesion}
Participating in collective decision-making can increase understanding of \textit{others} and learning about a topic's diverse beliefs and experiences. This can reduce inter-group prejudice \cite{pettigrew2006meta}, though sometimes it may backfire and increase polarization \cite{bail_cross_cutting}. We aimed to design a system encouraging perspective exploration while increasing mutual understanding and respect.

\subsubsection{Learning}
We aimed to support individual and group learning. Individually, participants should gain clearer understanding of policy issues. Collectively, they should learn from diverse perspectives, values, and experiences—helping people form more informed opinions and enabling stronger group decisions \cite{landemore2012many}.

\subsubsection{Scale of Participation}
In-person deliberation often increases process legitimacy and social cohesion but lacks scalability (often only 100-200 participants can join) and is expensive: assemblies can cost hundreds of thousands of dollars \cite{Involve_CitizensAssemblyCost_2024}. Our goal was to use LLMs to enable greater, cheaper participation while maintaining process legitimacy and social cohesion.

% 1. process legitimacy
%     Trust in outcomes
%     Understanding of decisions
%     Willingness to abide
%     Feeling heard
% 2. Social cohesion
%     Respect and rationality
%     Willingness to engage
%     Commonalities
% 3. Scale

\subsection{User Interface Design}
Our user interface had two parts: an AI interviewer where participants discussed their experiences on policy topics, and a visualization showing predicted positions on these issues based on the interviews. Below, we connect each component to our objectives.

% \begin{figure}
%     \centering
%     \includegraphics[width=1\linewidth]{figures/AI-Interviewer.png}
%     \caption{\added{AI-interviewer interface. Figure source: Park et al. (2024) supplementary materials, page 7 \cite{park2024generativeagentsimulations1000}}}
%     \label{fig:ai-interviewer}
% \end{figure}

\subsubsection{STEP 0: AI Interview}
Traditionally, collective decision-making has faced a trade-off between scale (how many people can participate) and depth (how much nuance their preferences can convey). Achieving mass participation, political equality, and deliberation at once is difficult \cite{fishkin_trilemma}. Historically, collecting and analyzing rich qualitative data at large scales has been nearly impossible.

LLMs are shifting this constraint by enabling collection and synthesis of nuanced qualitative data at scale \cite{li2023elicitinghumanpreferenceslanguage}. AI-driven interviews can gather deep insights and preferences from large numbers of participants efficiently ~\cite{chopra2023conducting, park2024generativeagentsimulations1000, wuttke2025ai} while adapting to their responses in real-time \cite{bachmann2025adaptive}. We leveraged this capability using a voice-based AI interviewer that conducted 45-minute interviews with participants about their backgrounds, life experiences, and policy beliefs.
We used interviews over surveys to elicit nuanced, personalized experiences while ensuring consistency through a deterministic script with predefined questions that limited the agent's own opinions \cite{park2024generativeagentsimulations1000}. This enabled scalable, uniform data collection of 90+ interviews in two days—infeasible with human interviewers \cite{park2024generativeagentsimulations1000}. 

The AI interviewer (Figure \ref{fig:steps}) was adapted from Park et al. (2024) \cite{park2024generativeagentsimulations1000}, prioritizing low latency and voice-to-voice interaction to simulate the feeling of talking to an interviewer. Participants simply spoke their answers; the system automatically detected completion, transcribed speech, and generated the next question vocally. We used interview data to predict participants' positions on policy proposals and highlight their relevant experiences.
Like a semi-structured interview, the LLM began with background questions about the interviewee’s life and then moved to three policy topics: minimum wage, race and gender in hiring, and domestic vs foreign hiring. We captured both beliefs and personal experiences grounding them: \textit{“Have you or someone close to you ever been impacted by immigration policy around work?''} and \textit{“How do you feel about companies prioritizing local over foreign applicants?''}. Check \autoref{app:ai-interviewer} for full list of questions and LLM prompts.

\subsubsection{Voting Visualization}
\begin{figure*}
    \centering
    \includegraphics[width=0.95\linewidth]{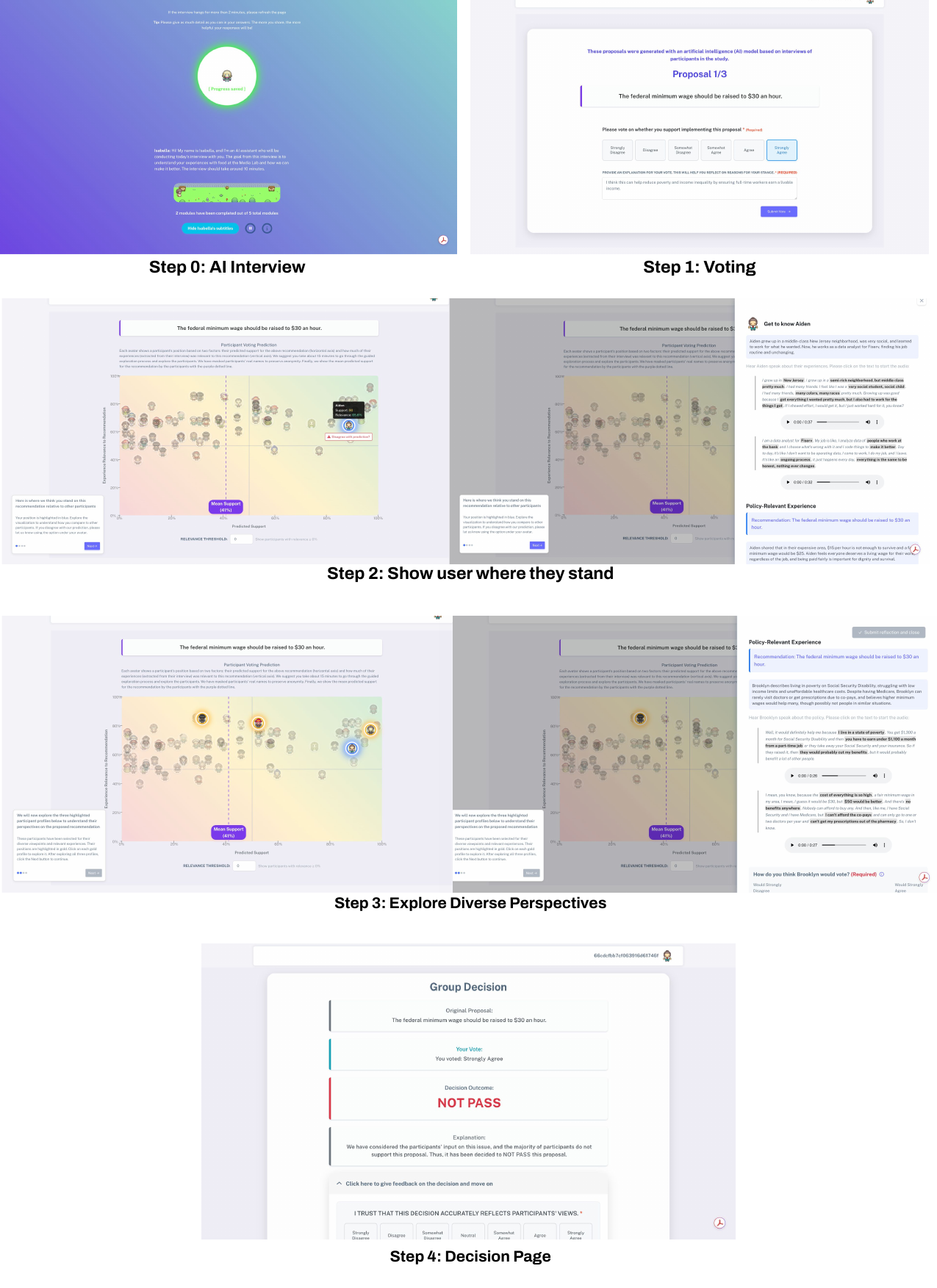}
    \caption{An overview of the different elements used in our study's process (Detailed version in \autoref{app:tool_screenshots})}
    \label{fig:steps}
\end{figure*}

After the AI interview, participants explored others' perspectives and experiences on each policy through a visualization interface. The steps (\autoref{fig:steps}) are outlined below.

\paragraph{STEP 1: Proposal Voting}

Participants first saw a proposal screen with a single policy statement. A transparency cue stated that \textit{``proposals were generated with an AI model based on interviews of participants in the study.''} Participants voted on a 6-point Likert scale (Strongly Disagree → Strongly Agree), avoiding a neutral midpoint to encourage directional commitment. They then provided a free-response justification to elicit reasoning, not just positions—aligning with deliberative principles \cite{fishkin_trilemma} and evidence that articulated reasons can reduce misperception and polarization \cite{puryear2024using}.

\paragraph{STEP 2: Showing the User Where They Stand: } 

After initial voting, participants saw a two-dimensional plot visualizing the crowd. Each avatar represents one interviewee, positioned by model derived predicted support (horizontal axis, 0–100\%) and experience relevance (vertical axis, how closely their interview concerned lived experiences salient to the proposal). We also distinguished stated opinions from concrete lived experiences (see section \ref{implementation:viz}). A purple dashed line shows mean support for transparency, akin to public polling breakdowns \cite{small2021polis, ovadya2023bridging}. The user's avatar and prediction appears highlighted in blue. An inline "Disagree with prediction?" control lets users correct their placement, building on contestability principles \cite{lyons2021conceptualising} to promote trust. Avatars with pseudonyms humanize the crowd while preserving anonymity, and a left–right color gradient visualizes the opinion landscape.
Clicking the blue avatar opens a panel with content from the user's interview (implementation detailed in section \ref{implementation:viz}): (i) a Life Story with an LLM-generated bio, clickable transcript excerpts, and synchronized audio from their interview self-introduction, preserving their voice; and (ii) Policy-Relevant Experience section surfacing proposal related experiences with transcript snippets and an LLM summary. This self-view increases transparency by showing which excerpts others will see and refreshing users' memory of what they shared. It clarifies how the system represents everyone while preserving authentic testimony.

\paragraph{STEP 3: Explore Diverse Perspectives: }

Participants then see three "featured" avatars in yellow, preselected to represent distinct stances across the predicted-support spectrum with high relevance to the proposal (selection algorithm in section \ref{implementation:viz}). A dialogue box prompts users to explore all three profiles before proceeding, countering confirmation bias by ensuring exposure to substantively diverse viewpoints. Clicking a featured avatar opens the profile view (voice clips + transcript snippets) with concrete, experience-based rationales \cite{kessler2023hearing, kubin2023reducing}. Built-in reflection scaffolds ask users to rate \textit{``How connected do you feel to Adrian?''} and \textit{``How do you think Adrian would vote on this policy?''}, with optional open-ended explanation boxes for perspective-taking. Users may also send optional ``connection requests'' to continue the conversation. After reviewing the three required profiles, participants can explore other avatars freely.
This balances scalability with in-depth exposure. Instead of skimming a list of pros and cons, participants \textit{hear} the reasons behind beliefs \cite{kessler2023hearing, kubin2023reducing}, helping them recognize key arguments and approach other profiles with greater empathy and openness.

\paragraph{STEP 4: Decision Page: }
After exploration, users see a final decision screen summarizing the policy outcome: reporting majority vote, resulting decision, and brief justification. An embedded feedback panel gathers post-decision reflections via Likert questions.

% —covering trust in the outcome, perceived process legitimacy, perceived common ground with others, views of others’ rationality, willingness to comply, and post-decision confidence in one’s own stance. This page closes the loop by making the tally explicit and immediately soliciting users’ evaluations of the process.

These four steps were repeated for each of the three proposals.

\begin{figure}
    \centering
    \includegraphics[width=1\linewidth]{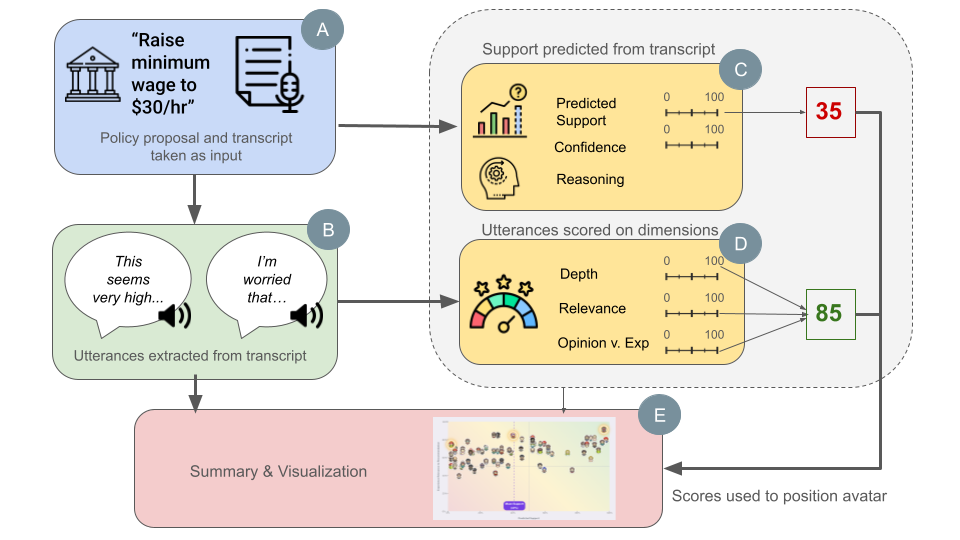}
    \caption{Procedure for curating data for the visualization}
    \label{fig:viz_preds}
\end{figure}

\subsection{Implementation}
The tool was built using the Django 4.2.5 web framework and hosted on AWS Elastic Beanstalk for easy deployment and scalability.

\subsubsection{AI Interviewer}
For the AI interviewer, adapted from \citet{park2024generativeagentsimulations1000}, we used OpenAI's Whisper to transcribe speech to text \cite{radford2023robust}, GPT-4o to generate responses \cite{hurst2024gpt}, and OpenAI's tts-1 to convert those back to speech \cite{openai_tts1}. The interview allocated an estimated time for participants' responses to each question, allowing the LLM to ask follow-ups before proceeding to the next scripted item.

\subsubsection{Voting Visualization}
\label{implementation:viz}
\paragraph{Participant Background}
Before engaging with policy-related experiences, we shared a brief summary of the individual’s background with 1-2 relevant transcript excerpts to help participants understand their story. See \autoref{app:viz-prompts} for LLM life prompts.

\paragraph{Policy Support and Relevance}

\autoref{fig:viz_preds} shows the steps described below. For the visualization, we wanted to estimate each participant's support and relevance of their experiences for every policy.
\textbf{Step A}: We hence began with the full interview transcript and the policy proposal. 
\textbf{Step B}: Next, GPT-4.1 extracted policy relevant utterances from the interview transcript, prioritizing in-depth personal experiences shown to bridge moral and political divides \cite{kubin, kessler2023hearing}.
\textbf{Step C}: GPT-4.1 was prompted with the transcript and proposal to rate predicted support (0–100), provide reasoning, and give a confidence score. Asking for reasoning before answers increases accuracy \cite{wei2023chainofthoughtpromptingelicitsreasoning}. We used predictions rather than pre-reported support because LLMs are increasingly used for policy preference estimation \cite{jarrett2025languageagentsdigitalrepresentatives, park2024generativeagentsimulations1000}; this also let us test prediction accuracy and examine participant reactions to seeing their own and others' predictions (most work tests accuracy alone, not responses to predictions in deployment). Though we didn't isolate this effect, we explored these perceptions in follow-up interviews.
We validated predictions against interviewee's initial stances, finding 82\% average accuracy across policies (\autoref{app:predicted-support}).
\textbf{Step D}: A separate prompt scored the extracted utterances from step B on depth, relevance, and the extent to which each reflected an opinion vs. a personal experience. These scores were averaged into a composite "relevance" score.
\textbf{Step E}: The support score (Step C) and relevance score (Step D) together determined each participant's avatar position in the visualization, as well as featured profiles selection. When a user clicks on an avatar, they will see and hear the utterances selected in Step B. See \autoref{app:viz-prompts} for the full set of LLM prompts.

\paragraph{Featured Profile Selection}
Since reviewing every profile was unrealistic, we guided participants to explore three profiles representing the full support spectrum. We grouped profiles into Low (0–33), Medium (33–66), and High (66–100) Support, then sampled one profile with relevance score $\geq 70$ from each category. This ensured participants encountered diverse, high-quality, relevant experiences without results being overly shaped by any single profile set.

% (\textbf{A}) We begin with the full interview transcript and the relevant policy. (\textbf{B}) We extract up to two utterances relevant to the policy that capture the person's beliefs or experiences. (\textbf{C}) From the full interview transcript, we predict the person's support from the policy. (\textbf{D}) We score the utterance on depth, relevance, and whether they capture opinions versus experiences and then combine these into a single quality score. (\textbf{E}) We use the predicted support and quality score to position the avatar. When a user clicks on the avatar, they will see and hear the utterances selected in \textbf{B}.

\section{Methods} 

\subsection{Study Context and Participants}
Our study was IRB-approved with informed consent obtained from all participants. We recruited from Prolific \cite{prolific}, requiring participants to commit one continuous hour, have desktop access, consent to audio sharing with other participants, and agree to return for a second phase one week later. We filtered for participants with $>95\%$ approval ratings on the Prolific platform, currently living in the U.S., and at least one year of work experience. We also used Prolific's sampling settings to recruit a politically balanced mix of liberal, moderate, and conservative participants.

In phase two, we invited participants to \textit{``Vote on proposals for how to make workers' lives better!''} Due to our \(2 \times 2\) factorial design, we re-invited phase one participants \textit{and} recruited new participants who hadn't done the AI interview, applying the same filters for recruitment as phase one. See \autoref{app:demographics} for demographic breakdown. Participants received \$10/hour.

\subsection{Experimental Design}
We ran a randomized controlled trial ($n=181$) with four conditions: 

[\textbf{Condition A}, $n=42$]: Took AI interview and saw voting visualization. Completed all steps from \autoref{fig:steps}.

[\textbf{Condition B}, $n=48$]: Took AI interview but did not see voting visualization. Completed steps 0, 1 and 4 from \autoref{fig:steps}.

[\textbf{Condition C}, $n=44$]: No AI interview but saw voting visualization. Completed steps 1, 2, 3, and 4 from \autoref{fig:steps}.

[\textbf{Condition D}, $n=47$]: No AI interview and did not see voting visualization. Completed only steps 1 and 4 from \autoref{fig:steps}.

% To clarify, participants in groups A and B completed the AI interview and were subsequently invited to the second phase of the study; however, only group A interacted with the visualization tool, while group B did not. In phase two, we also recruited new participants who had not taken the AI interview, assigning them to groups C and D, where group C viewed the visualization and group D did not. 
This \(2 \times 2\) factorial design enabled us to examine the effects of both the AI interview and the visualization.

\subsection{Measurement Framework}
\label{sec:measured_vars}

Our measurement framework has three tiers: domains (high-level goals), concepts (specific aspects within domains), and items (individual questions). The three domains are tied directly to our system objectives: \textit{process legitimacy}, \textit{social cohesion}, and \textit{stances and learning}. Each domain contained multiple concepts assessed by one or more survey items, which we describe below. 
% See \autoref{tab:domain_conc} for concepts within each domain with descriptions. 
Full survey items are in \autoref{app:reg_results} and item reliability within each concept is in \autoref{app:survey_reliability}.

\subsubsection{Process Legitimacy}
These outcomes capture participants’ perceptions of the fairness and transparency of the decision-making process. They include concepts of 1) trust in the process, 2) willingness to comply with resulting decisions, 3) understanding of how decisions were reached, and 4) the extent to which participants felt 
their voices were heard. 

\subsubsection{Social Cohesion}
These outcomes reflect the degree to which the process fostered 1) a sense of connection among participants. They include perceptions of  2) rationality and 3) mutual respect, 4) gained insight into diverse viewpoints different from their own, 5) ability to understand the origins of others’ beliefs, 6) respect for pluralism and 7) whether they recognized shared commonalities and felt more 8) curious and 9) willing to engage with other participants.

\subsubsection{Stances and Learning}
We measured 1) participants’ stances on policies and 2) their confidence in those stances before and after the process to 3) record changes, and 4) whether they learned something about each of the topics through the process.

\section{Study Procedure} 
We outline the stages of the study here, and put in parentheses which participant conditions participated in each stage.
% For participants that took the AI interview, there were two main phases. In the first phase, the participants took a pre-survey and then the AI interview. They then returned a week later to vote on the proposals, interact with the tool, and take a post survey. The participants that did not take the AI interview only participated in phase two, where they took the pre-survey, voted on the proposals, interacted with the tool, and took the post survey. We outline the phases in detail below.
\subsection{Pre-survey (A, B, C, D)}
\label{sec:pre-survey}
Participants were routed from Prolific to Qualtrics for a pre-survey capturing demographics, policy stances (minimum wage, race/gender in hiring, domestic vs. foreign hiring), and respect for pluralism.

\subsection{Phase One: AI Interview (A, B)}
Participants then logged into the tool platform using their Prolific ID, created their avatar, and took the voice-driven AI interview.

\subsection{Phase Two: Voting and Tool Interaction}
Participants voted on three different proposals shown in random order: ``Race and gender should be used in hiring decisions to combat inequality in the workplace'', ``The federal minimum wage should be raised to $\$30$ an hour'', and ``Companies should strongly prioritize hiring domestically before considering foreign applicants.''
\subsubsection{Policy Voting (A, B, C, D)}
Participants first were shown a screen where they were asked to vote on one of the proposals via a six point Likert scale. This screen is shown under step 1 in \autoref{fig:steps}.

\subsubsection{Visualization Interaction (A, C)}
\label{sec:vis-interaction}
Participants used the visualization for a guided exploration of the other participants' experiences on the topic. After exploring three featured profiles, they could explore other profiles if they desired. See step 3 in \autoref{fig:steps}.

\subsubsection{Decision Page (A, B, C, D)}
Participants then saw a simulated policy decision page (pass/not pass) and answered questions about their trust in the decision, willingness to abide by it, and understanding of others' stances.
A key design choice was participants always saw the final decision \textit{go against} them. Since individuals scrutinize fairness most when outcomes are unfavorable \cite{brockner1996integrative}, this let us assess whether the system could foster trust, acceptance, and cohesion under disagreement, aligning with the democratic importance of "losers' consent" \cite{losers_consent}. To align outcomes with the visualization, we precomputed biased profile sets so mean predicted support matched the displayed result during \ref{sec:vis-interaction}. For example, raising the minimum wage supporters saw profiles with aggregate support opposing it, with the visualization showing the policy didn't pass; opponents saw the inverse distribution. This ensured comparable experiences across conditions and controlled for variation in participant stance. Full profile sampling details are in \autoref{app:samp_prof}.

\subsection{Post-survey}
\label{sec:post-survey}
Finally, participants returned to Qualtrics for the post-survey, which repeated the pre-survey stance and pluralism questions and added questions on process trust and social cohesion (from \autoref{sec:measured_vars}).

\subsection{Data Collection}
\subsubsection{Surveys}
Participants completed Qualtrics surveys in two places: a pre-survey at the beginning (\autoref{sec:pre-survey}) and a post-survey at the end (\autoref{sec:post-survey}). Questions primarily used seven-point Likert scales with optional text boxes for elaboration.

\subsubsection{Interviews}
% To further probe how participants across conditions experienced the process and outcomes, the two lead authors conducted 22 semi-structured Zoom interviews (20–30 minutes each). Participants were recruited via a post-survey opt-in and received an additional \$10. We interviewed eight people from \textbf{Condition A}, six from \textbf{Condition B}, four from \textbf{Condition C}, and three from \textbf{Condition D}—stopping when we reached thematic saturation. Interviews were audio-recorded (with consent) and followed a protocol that began with a brief warm-up about the overall study experience. 
% We explored how participants experienced the process end-to-end—whether they trusted it, felt represented, and understood how their interview informed proposals; their willingness to accept outcomes (including when they disagreed) and what would influence compliance; feelings of connection and empathy toward others and perceived commonalities; views on AI’s role in fairness, quality, and inclusion and where human oversight should sit; comfort and openness with the AI interviewer; moments of learning or perspective-taking; and, finally, what changes or assurances would make them trust and use such a system in workplaces or civic contexts.

To examine experiences across conditions, the two lead authors conducted 22 semi-structured Zoom interviews (20–30 mins each), recruited via post-survey opt-in (\$10 incentive). We interviewed eight from Condition A, six from Condition B, four from Condition C, and three from Condition D till we reached thematic saturation.
We explored participants' trust, representation in process, outcome acceptance (especially under disagreement), empathy and connection with others, perceptions of AI's role in fairness and quality, comfort with the AI interviewer, perspective-taking moments, and factors influencing real-world adoption.

% \subsubsection{Interaction Tracking and Analytics}

\subsection{Quantitative Data Analysis}
% \subsubsection{Power Analysis}
\subsubsection{Creating Concept Measures}
For each concept in \autoref{sec:measured_vars}, we asked multiple survey questions to ensure robustness to phrasing effects \cite{Choi2005Catalog}. To aggregate into a single concept score, we normalized responses to each question (across all conditions), then averaged the normalized responses within each concept to create a composite concept score for each participant.

% This approach helped to smooth out variability due to question phrasing, a kn

% but also makes the coefficients more interpretable, as they can now be understood in terms of the effect of the treatment measured in standard deviations of the underlying concept.
\subsubsection{Estimating Treatment Effects}
We used a standard regression framework for our \(2 \times 2\) factorial design. For each concept, we used the following model, regressed the concept measure on binary indicators for AI interview completion and visualization viewing, the interaction between the two, and control variables. We controlled for demographics (age, gender, income, education, ethnicity, employment status, political orientation), as well as initial policy support, average stance confidence, and respect for pluralism.

\begin{equation}
Y_i = \beta_0 
      + \beta_1 \,\text{AI}_i
      + \beta_2 \,\text{Viz}_i
      + \beta_3 \,(\text{AI}_i \times \text{Viz}_i)
      + \boldsymbol{\gamma}^\top \mathbf{X}_i
\end{equation}

\noindent
In this specification, \(Y_i\) denotes the concept measure for participant \(i\). Binary indicators \(\text{AI}_i\) equals 1 if the participant completed an AI interview, 0 otherwise; \(\text{Viz}_i\) equals 1 if shown the visualization, 0 otherwise. The interaction term \(\text{AI}_i \times \text{Viz}_i\) captures joint treatment effects. The vector \(\mathbf{X}_i\) comprises the control variables described above with coefficient vector \(\boldsymbol{\gamma}\) measuring their effects.

\subsubsection{Analyzing Item-level Responses}
While we focus on concept-level effects in the main manuscript, we also analyzed effects on individual survey items using the same regression but with each item as a single outcome $Y_i$. To account for multiple testing within each concept, we applied the Benjamini–Hochberg procedure \cite{bh}.

% \subsubsection{\added{Analyzing Audio Telemetry MOVE TO APPENDIX}}
% \added{
% Prior work has shown that listening to users experiences via audio segments can lead to more positive social evaluations of others, compared to reading text segments \cite{kessler2023hearing} and emphasize the emotional impact of the depicted data \cite{elli2020tied}. 
% Although we did not conduct a separate ablation study to isolate the effects of listening to the audio recordings on our tool (as this was outside our primary research objectives), we collected telemetry data on audio player usage and the number of participant profiles each user explored. 
% To examine the relationship between this behavioral telemetry and post-intervention concept measures, we conducted additional regression analyses with telemetry metrics as predictors and the same concept measures as outcomes. Telemetry variables included average profiles explored and average audio listening duration, aggregated across three recommendations. All telemetry predictors were standardized prior to analysis. We once again controlled for demographic variables, pre-survey baseline measures, and interview condition (whether they gave the AI interview prior to tool use or not). 
% }

\subsection{Qualitative Interview Analysis}
We transcribed and cleaned all 22 interviews, then conducted inductive thematic analysis \cite{charmaz2008grounded, corbin1990grounded}. One of the lead authors read each transcript for analytic notes; a second pass generated preliminary codes, refined through discussion with other authors. Related quotes were clustered into high-level categories, themes, and definitions; these were revised until consensus and thematic saturation. We then examined thematic variation across experimental conditions, consulting the broader team to validate interpretations. This qualitative analysis contextualized the quantitative survey findings.

\section{Quantitative Results}
\begin{figure}
    \centering
    \includegraphics[width=1\linewidth]{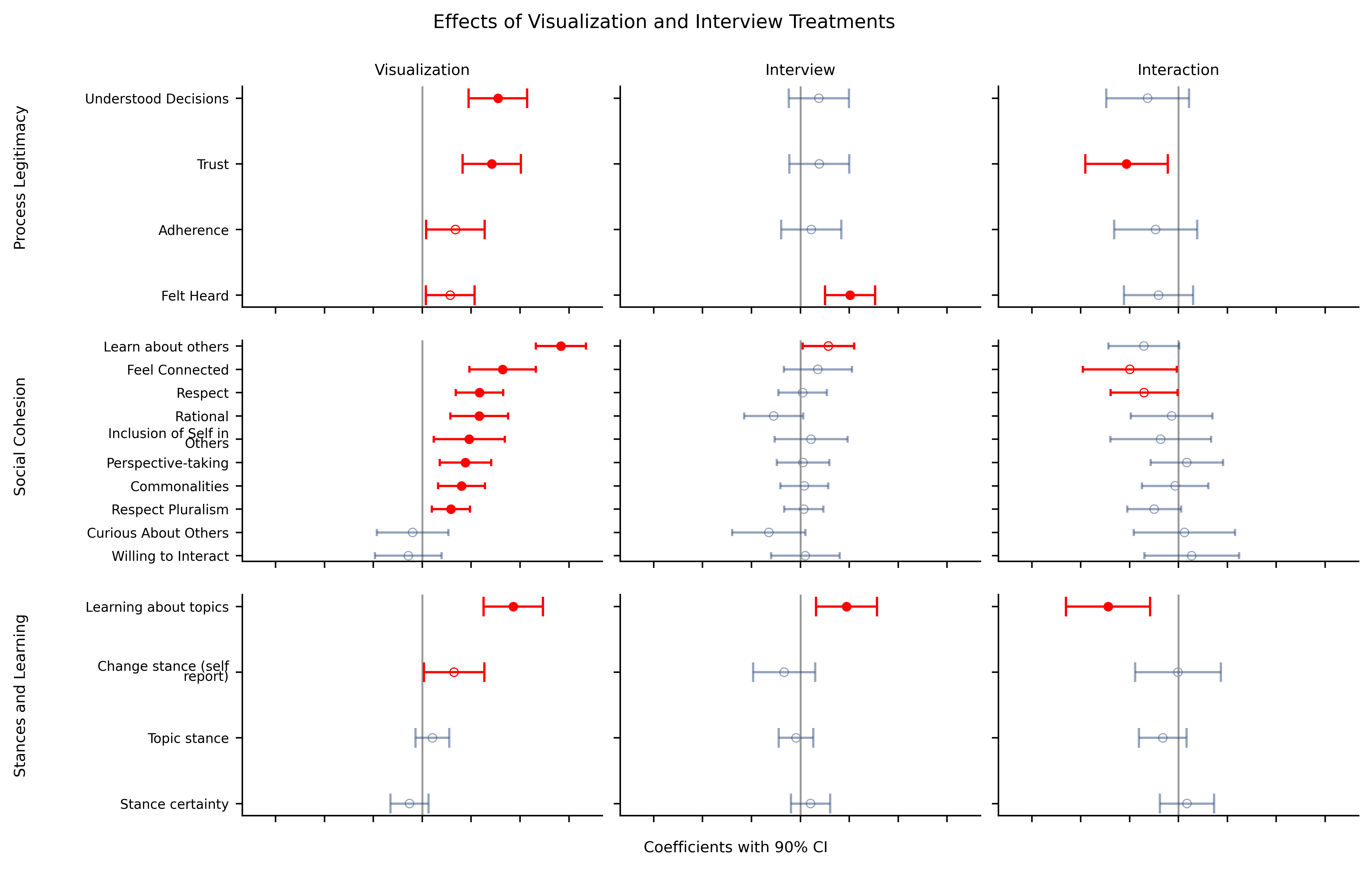}
    \caption{
    Coefficients for visualization, AI interview, and their interaction (90\% CI). Red: \textit{p < 0.1}; solid markers: \textit{p < 0.05}.
    }
    \label{fig:forest}
\end{figure}

Results for concepts are in \autoref{fig:forest}. 
We report the coefficient of the treatment of interest as well as the $90\%$ confidence interval and p-value. Since all outcomes were standardized, coefficients can be interpreted as effect sizes in standard deviation units.

\subsection{Visualization}
The visualization enhanced process legitimacy and social cohesion. Participants reported better understanding of decisions ($\beta = 0.77$, CI = [0.47, 1.07], $p < .001$, $R^2_{adj} = 0.18$) and greater trust in outcomes ($\beta = 0.71$, CI = [0.41, 1.01], $p < .001$, $R^2_{adj} = 0.13$). They showed higher willingness to adhere ($\beta = 0.34$, CI = [0.04, 0.64], $p = .06$, $R^2_{adj} = 0.23$) and stronger sense that their input mattered ($\beta = 0.29$, CI = [0.04, 0.54], $p = .06$, $R^2_{adj} = 0.21$), though these effects were smaller. 
The visualization improved most social cohesion measures, except curiosity about others and willingness to interact with them, we discuss this further in the qualitative section.
Participants viewing the visualization reported learning more about the topics ($\beta = 0.93$, CI = [0.63, 1.23], $p < .001$, $R^2_{adj} = 0.28$) and self-reported shifts in stance ($\beta = 0.32$, CI = [0.02, 0.63], $p = .08$, $R^2_{adj} = 0.50$). 
However, we found no meaningful visualization effects on actual policy support ($\beta = 0.10$, CI = [$-0.07$, 0.28], $p = .32$, $R^2_{adj} = 0.50$) or stance certainty ($\beta = -0.13$, CI = [$-0.33$, 0.07], $p = .27$, $R^2_{adj} = 0.33$). 

Though outside our RCT RQs, we report: Conditions A and C explored an average of 4.04 profiles and listened to 358.34 seconds (6 minutes) of audio. \autoref{app:audio_results} shows the regression analysis we conducted using these telemetry variables.
While audio and profile exploration had limited effects overall, listening to audio increased social connectedness ($\beta = 0.282$, CI = [0.043, 0.521], $p < 0.1$) and self-reported stance changes ($\beta = 0.399$, CI = [0.114, 0.684], $p < 0.05$), corroborated by qualitative interviews.

% \added{Participants who had access to the visualization (Conditions A and C) explored an average of 4.04 profiles and listened to 358.34 seconds (6 minutes) of audio on average. Figure \ref{fig:audio_regression} [ADD TO APPENDIX] shows the regression analysis we conducted using these telemetry variables. The audio and profiles explored did not significantly influence most of our concepts. However, listening to the audio did correlate with increase in feelings of social connectedness with other participants of the study ($\beta = 0.282$, CI = [0.043, 0.521], $p < 0.1$). This was also corroborated by our qualitative interview results (see next section). Interestingly, we also saw a positive correlation between audio usage and changes in self-reported stances on the topics ($\beta = 0.399$, CI = [0.114, 0.684], $p < 0.05$). We want to emphasize that these trends are correlational rather than causal (e.g. it could be that certain types of respondents are more likely to listen to profiles and also are likely to report higher measures of connection), but we reported them to shed light on the relationship between engagement with the profiles and our outcome measures.

\subsection{AI Interview}
The interview did not significantly influence most concepts. However, participants who took the AI interview felt more heard ($\beta = 0.51$, CI = [0.25, 0.76], $p < .001$, $R^2_{adj} = 0.21$) and reported learning more about the topic ($\beta = 0.47$, CI = [0.16, 0.78], $p = .01$, $R^2_{adj} = 0.28$) and about others ($\beta = 0.29$, CI = [0.02, 0.55], $p = .07$, $R^2_{adj} = 0.47$)—despite the interview eliciting their own beliefs rather than providing information about the topic or others. Reflection on personal beliefs through interviews may have increased self-reported learning. Results for individual survey items are in \autoref{app:reg_results}.

\subsection{Interaction between Visualization and AI Interview}
While the visualization enhanced trust and learning overall, effects were strongest for participants who did \textit{not} take the AI interview ($\beta = -0.53$, CI = [-0.95, -0.11], $p = .04$, $R^2_{adj} = 0.13$) ($\beta = -0.72$, CI = [-1.15, -0.29], $p = .01$, $R^2_{adj} = 0.28$).
Similar reductions appeared for respect ($\beta = -0.35$, CI = [-0.70, -0.01], $p = .09$, $R^2_{adj} = 0.41$) and connection ($\beta = -0.50$, CI = [-0.98, -0.02], $p = .09$, $R^2_{adj} = 0.22$). 
Interestingly, participants who didn't take the interview showed stronger positive effects from the visualization than those that did (though the overall visualization effect was positive). Perhaps, those interviewed felt more invested in the process and reacted more negatively when decisions went against them.

\section{Qualitative Results}

\autoref{app:themes} outlines interview themes and their definitions. Participant ID's have been appended with condition type for comparison.

% \begin{itemize}
%     \item A:	Took AI interview AND saw voting visualization
%     \item B:	Took AI interview but did NOT see voting visualization
%     \item C:	NO AI interview but saw voting visualization
%     \item D:	NO AI interview and did NOT see voting visualization
% \end{itemize}

\subsection{Social Cohesion}

\subsubsection{Social Connection with Other Participants}
Participants' social connectedness, empathy, and openness to engagement varied by how much of people's life stories they could ``hear''.

In \textbf{Condition A}, audio functioned as social glue: \textit{"Being able to see their background and actually hear their voice made them seem like actual real people behind opinions"} (A3). Voice conveyed affect and credibility that text couldn't: \textit{"I could really understand their emotions a lot more when I heard audio...it made me relate"} (A4), deepening empathy toward life histories that \textit{"lent a lot more credence"} to stances (A4). Participants found commonalities around shared identities or pro–working-class values. However, without two-way exchange, \textit{"I don't feel any connection to a stranger just by hearing a story"} (A2); backstories clarified reasoning but rarely shifted positions: \textit{"knowing their background helped me understand why they feel that way, but I still disagree"} (A4).

For those in \textbf{Condition B}, social connection was notably thinner. Many described the experience as opaque or \textit{“relationship-less”}: with no visibility into others’ viewpoints, they felt disengaged—\textit{“I have no insight into what they’re doing”} (B3). Still, some expressed a strong desire for engagement features: B6 asked for \textit{“an ability to engage with the other voters… even a brief session,”} or at minimum see \textit{“profiles… occupation, education, age, gender.”} 

% For \textbf{Condition C} hearing others again became a catalyst for connection, empathy, and curiosity similar to \textbf{Condition A}. C2 appreciated the combination of seeing stances and hearing reasons behind them. Exposure to others prompted a desire to follow up and have sync conversations. However, not everyone wanted to reach out—C4 feared confrontation with those opposed to their views wouldn’t \textit{“end… positive(ly) for either of us.”} Even so, backstories helped many rethink assumptions. C1 reconsidered the “hire locals first” stance after hearing a healthcare perspective about labor shortages. Barriers remained when demographics felt distant—C4 had trouble relating to some profiles when there were significant differences in age and geographical background. Participants also asked for more explicit reasoning and sources: C4 suggested prompts like \textit{“what resources did you use to form this view?” }to avoid relying on inference from biography alone or mistaking monotone delivery for neutrality.

In \textbf{Condition C}, hearing others again catalyzed connection, empathy, and curiosity, mirroring \textbf{Condition A}. Exposure to others' stances and reasoning prompted desires for follow-up conversations, though not universally: C4 feared confrontation with opposing views wouldn't \textit{"end… positive(ly) for either of us."} Backstories also helped participants rethink assumptions; C1, for instance, reconsidered their "hire locals first" stance after hearing a healthcare worker describe labor shortages. C4 suggested adding prompts like \textit{"what resources did you use to form this view?"} to avoid over-relying on inferences from biography alone.

% Social connection was the most constrained in \textbf{Condition D} when participants had neither interview nor visualization. Without the ability to hear or see others, people called the experience \textit{“so impersonal” }(D3) and worried about whether the participants were even real—\textit{“AI’s gotten so smart, you don’t know.”} Still, several \textit{wanted} to engage. D3 explained flatly that empathy requires \textit{knowing}: \textit{“if this was somebody who had jumped through hoops… that would definitely make me… sympathize… but not knowing.”} Across conditions B and D, then, the main barrier to social connectedness was a lack of exposure to others’ voices, stories, and reasoning, compounded by no affordances for conversation. \added{However, we acknowledge that this condition, though realistic in the sense that many decisions go against one's beliefs with no input or exploration of others' beliefs, may have been artificially weak for social cohesion in this condition since they had no input to the process or opportunity to engage with other perspectives.}

Social connection was most constrained in \textbf{Condition D}, where participants had no audio or visualization. Without any way to see others, participants called the experience \textit{"so impersonal"} (D3) and questioned whether other participants were even real: \textit{"AI's gotten so smart, you don't know."}  Yet several still wanted to engage: D3 put it plainly that empathy requires knowing: \textit{"if this was somebody who had jumped through hoops… that would definitely make me… sympathize… but not knowing."} 
% Across Conditions B and D, then, the primary barrier to social connectedness was lack of exposure to others' voices and stories compounded by no affordances for conversation.
Condition D—while realistic in that many decisions go against one's beliefs without any input—may have been artificially weak for social cohesion, given participants had no opportunity to engage with or influence the process.

% Taken together, the presence of audio + transcripts of participant experiences (conditions A, C) consistently increased perceived empathy and curiosity—participants said hearing real voices turned \textit{“text on a screen”} into \textit{“actual real people”} (A3), revealed conviction, and invited rethinking. In their absence (Conditions B and D), participants defaulted to skepticism. 

\subsubsection{Rationality, Understanding, and Respect for Others' Views}
Participants’ perceptions of the rationality behind diverse opposing views varied by condition, producing different levels of respect.

% In \textbf{Condition A}, participants articulated concrete criteria for what would make others’ views feel \textit{rational} and worthy of deference: engagement with counter-arguments, explicit logic, and connections between personal experience and policy implications. A2 was explicit that profiles that \textit{“just told me what [they] think”} without addressing opposing points—one-off anecdotes (\textit{“sob stories”}) were not sufficient to justify broad policy choices. A1 similarly differentiated between \textit{“me-centered”} claims (e.g., minimum wage framed only as personal gain) and population-level reasoning (e.g., a nurse’s account of staffing shortages), which she considered more valid even if it did not fully change her stance. Respect was thus often conditional on perceived depth, counter-positioning, and relevance. 

In \textbf{Condition A}, participants articulated concrete criteria for what made others' views feel rational and worthy of deference: engagement with counter-arguments, explicit logic, and links between personal experience and policy. A2 dismissed profiles that \textit{"just told me what [they] think"} without addressing opposing points; one-off \textit{"sob stories"} were insufficient to justify broad policy choices. A1 similarly distinguished \textit{"me-centered"} claims (e.g., minimum wage as personal gain) from population-level reasoning (e.g., a nurse's account of staffing shortages), treating the latter as more valid even without fully shifting their stance. Respect, then, was conditional on perceived depth, counter-positioning, and relevance.

% \textbf{Condition B} sharpened a distinct point: without access to \textit{sufficient arguments}, participants found it hard to judge others’ rationality and thus to offer informed respect. B3 noted that both the interview and voting phases \textit{“flattened nuance”} into binary outcomes, leaving people to be judged by a number rather than their reasoning. B5 explicitly wanted to see \textit{“the arguments for their side”} when decisions went the other way; B6 described peers as generally well-meaning but \textit{“not fully thought out” }on downstream effects (e.g., small-business impacts of a \$30 minimum wage) of policy decisions. Respect was thus affirmed as a norm, yet judgments of others’ rationality were constrained and often hypothetical and generic in the absence of concrete justifications. 

In \textbf{Condition B}, without access to others' reasoning, participants struggled to assess rationality and offer informed respect. B3 noted that both the interview and voting phases \textit{"flattened nuance"} into binary outcomes, reducing people to numbers; B5 wanted to see \textit{"the arguments for their side"} when decisions diverged; and B6 described peers well-meaning but \textit{"not fully thought out"} on downstream effects such as the small-business impacts of a \$30 minimum wage. Respect was affirmed as a norm, but judgments of rationality remained hypothetical without concrete justifications.

% In \textbf{Condition C}, participants anchored respect in \textit{how} claims were justified. Respect rose when decisions were \textit{traceable}—being able to see individuals' support and link them to concrete constraints made others’ positions legible and, thus, reasonable (C2). Participants here explicitly \textit{evaluated argument quality}: they flagged when peers conflated concepts (e.g., treating inflation and minimum wage as the same), or leaned on dated gender norms (C4). 

In \textbf{Condition C}, respect was anchored in \textit{how} claims were justified. It rose when decisions were \textit{traceable}: linking individuals' stances to concrete constraints made positions legible and reasonable (C2). Other participants flagged when peers conflated concepts (e.g., treating inflation and minimum wage as the same) or leaned on dated gender norms (C4).

% Participants in \textbf{Condition D} adopted an “accepting but skeptical” stance—acknowledging their view is not fact, staying open to being in the out-group, and \textit{“keeping [their] mind open to all the possibilities”} (D1). Others described an immediate disbelief when outcomes contradicted expectations, which shifted into a request for transparent justification—\textit{“I would want to know their reasoning… it might be a legitimate reason… it might… persuade me” }(D2). Because \textit{“everybody was a blank table,”} participants didn’t presume to know others’ stories and thus “left room for doubt,” treating opposing views as valid given  different life circumstances, even when they did not get to see it (D3).

In \textbf{Condition D}, participants adopted an accepting but skeptical stance. When outcomes contradicted expectations, initial disbelief shifted into a desire for justification: \textit{"I would want to know their reasoning…it might be a legitimate reason and persuade me"} (D2). Because \textit{"everybody was a blank table,"} they didn't presume to know others' stories, leaving room for doubt and treating opposing views as potentially valid given unseen life circumstances (D3).

% We will now explore how the above feelings of social cohesion and rationality impacted views on process fairness.

\subsection{Process Legitimacy}

\subsubsection{Participant Voice in the Process}

% A central concern was whether participants felt their voice was genuinely acknowledged. 
While all participants engaged in interviews, voting, or both, feeling "heard" varied markedly depending on the mechanisms available to their condition.

% \textbf{Condition A} offered both expressive and visible outlets for participation, but participants diverged on whether these mechanisms adequately conveyed their voices. Some expressed strong affirmation, describing the process as \textit{“really fair”} and appreciating that\textit{ “other people can vote whether or not they agree with my opinion, and then, I can vote on their opinion”} (A4). This reciprocity gave several participants confidence that their voice was part of a collective exchange. Yet some others were left uncertain, emphasizing the lack of feedback on whether \textit{their} specific avatar was “featured to at least a few participants (given that this choice was made by the system). For some, this translated into reflections on real-world democratic frustration—\textit{“you spend so much time and energy reading about different policies and then you go vote… and then, you know, no matter what… it felt like none of it even mattered”} (A4)—suggesting that perceptions of voice in the process were inseparable from larger narratives of political efficacy.

\textbf{Condition A} offered both expressive and visible outlets for participation, but participants diverged on whether these adequately conveyed their voices. Some described the process as \textit{"really fair"}, valuing the reciprocity of mutual opinion voting (A4). Others remained uncertain, troubled by the lack of feedback on whether their specific avatar had actually been seen by atleast some participants. This uncertainty bled into broader democratic frustration: \textit{"you spend so much time and energy reading about different policies and then you go vote…and then, you know, no matter what…it felt like none of it even mattered"} (A4), suggesting perceptions of voice were inseparable from larger narratives of political efficacy.

% In \textbf{Condition B}, participants often described the interview as personally validating: \textit{“I did feel heard… I really did”} (B2). Yet B3 stressed lingering ambiguity around voting and decision processes without the visualization, noting that limited transparency left them \textit{“between neutral and frustrated.”} Several tied their sense of voice to whether outcomes aligned with their preferences: \textit{“When it went my way… they must have listened. But if it failed… I wouldn’t feel that I was heard”} (B5). Others regretted the inability to exchange experiences, especially on sensitive topics like affirmative action, where personal narratives felt sidelined without dialogue (B6). Thus, interviews offered individual affirmation but, without collective transparency or deliberation, many remained ambivalent.

In \textbf{Condition B}, the interview was personally validating: \textit{"I did feel heard…I really did"} (B2) yet broader ambiguity persisted. B3 noted however that limited transparency around voting left them \textit{"between neutral and frustrated,"} and B5 tied their sense of voice directly to outcomes: \textit{"When it went my way…they must have listened. But if it failed… I wouldn't feel that I was heard."} Others regretted the inability to exchange experiences, particularly on sensitive topics like affirmative action where personal narratives felt sidelined without dialogue (B6). 
% Interviews thus offered individual affirmation but, without collective transparency or deliberation, many remained ambivalent.

% By contrast, \textbf{Condition C} foregrounded voting as the primary mechanism of voice. Several participants appreciated the direct equivalence of their input—\textit{“my vote counted just as much as everyone else”} (C1)—and the ability to explore others’ statements: \textit{“It made me realize… we all had a voice, we all had ability to make a choice, a decision” }(C2). However, some felt disadvantaged compared to those who had given interviews: \textit{“The only thing I was able to do was to vote yes or no… but most importantly, they could speak. And I think with them being able to speak, that gave meat to their vote”} (C2). One participant suggested design changes to strengthen voice, such as options to modify or reform aspects of proposals rather than being restricted to simple yes/no votes: \textit{“If there was… checkboxes of what you would like to see move forward with it… that would make it more apt to everybody” }(C4).

In \textbf{Condition C}, voting was the primary mechanism of voice. Participants appreciated that \textit{"my vote counted just as much as everyone else"} (C1) and the collective agency: \textit{"we all had a voice, we all had ability to make a choice, a decision"} (C2). Yet some felt asymmetrically disadvantaged relative to those who had given interviews: \textit{"The only thing I was able to do was to vote yes or no. But I think with them being able to speak gave meat to their vote"} (C2). Some also pushed for options to reform aspects of proposals: \textit{"If there was checkboxes of what you would like to see move forward with it that would make it more apt to everybody"} (C4).

% Finally, \textbf{Condition D} produced the sharpest disillusionment. Without expressive input or visible aggregation, participants described the process as hollow. As one participant put it bluntly: \textit{“Wow. I really didn’t [feel heard]. I was like, okay, this is some bull”} (D2). The absence of both interviews and visualization left little basis for perceiving one’s voice as recognized, making this condition the most exclusionary.

In \textbf{Condition D}, due to the absence of both expressive input and visible aggregation, participants described the process as hollow: \textit{"Wow. I really didn't [feel heard]. I was like, okay, this is some bull"} (D2)—making this the most exclusionary condition.

Participants thus equated feeling heard with opportunities for expression and transparency (seeing how contributions were used). 
% The results underscore that deliberative tools must go beyond capturing input—they must also provide participants with tangible signals of recognition, reciprocity, and influence.

\subsubsection{Process Clarity and Explanations}
Participants evaluated legitimacy through two lenses: \textit{can I see how this decision was produced?} and \textit{does what I see feel fair, even when it goes against me?} 
% Access to interviews and/or the voting visualization shifted those judgments in distinct ways.

% \textbf{Condition A:} Several participants framed the process as recognizable democratic choice and trusted the outcome on that basis:\textit{ “basically just voting… the decision was what the majority… wanted… fair is fair… it is what it is”} (A3). Others valued having both a place to speak and to observe aggregate results (A5). Yet even with both components, cracks appeared when outcomes conflicted with personal expectations. Owing to their previous experiences on Prolific, some speculated that the vote percentages \textit{“were not real,”} before concluding that \textit{“hearing myself” }convinced them real participant profiles were shown to them (A7). Others wanted more procedural transparency and explanations: guarantees that each profile would be seen by a minimum number of voters (A1), short rationales for pass/fail (\textit{“a sentence or two about… why it didn’t pass,”} A8), and even avenues to revise or reform proposals (\textit{“we could have changed it to make it pass,”} A8).   Participants were thus willing to accept majority rule, but still called for assurance mechanisms to ensure legitimacy.

In \textbf{Condition A}, several participants trusted the process as recognizable democratic choice: \textit{"basically just voting…the decision was what the majority wanted…fair it is what it is"} (A3). Yet cracks appeared when outcomes conflicted with expectations. Some speculated that vote percentages \textit{"were not real,"} before concluding that \textit{"hearing myself"} convinced them genuine participant profiles had been shown (A7). Others called for procedural transparency to shore up legitimacy: guarantees that each profile would reach a minimum number of voters (A1), \textit{"a sentence or two about why it didn't pass"} (A8), and avenues to revise proposals. 
% Participants were thus willing to accept majority rule, but called for assurance mechanisms to shore up legitimacy.

% \textbf{Condition B:} Without the visualization, many described the system as a \textit{``black box''}, warning that a flat \textit{“yes/no”} without reasons invites claims that \textit{“it’s rigged”} and can \textit{“manipulate”} reactions (B1). Participants wanted to know \textit{“how my answers would affect everything,”} \textit{“who said yes, who said no,” }and to \textit{“chat back and forth”} to understand others’ reasoning (B3). Some simply rejected the outcomes as illegitimate: \textit{“I didn’t view them as legitimate. I really thought that none of them were good responses”} (B4), asking for counts, counterarguments, and more factual information accompanying the decision (B4). Others noted outcome-dependent trust: \textit{“when it passed… great; when it fails… I question legitimacy,”} and called for vetting, visible tally counts, and even a\textit{ “governing body” }or\textit{ “trustworthy seal of approval”} to ground the process (B5). Even when participants trusted the results, they wanted space to explain why their own views diverged from the final outcome (B6). 

In \textbf{Condition B}, without visualization, many described the system as a \textit{"black box"} that \textit{"it's rigged"} and can \textit{"manipulate"} reactions (B1). They wanted to know \textit{"how my answers would affect everything,"} \textit{"who said yes, who said no,"} and to \textit{"chat back and forth"} to understand others' reasoning (B3). Some rejected decisions as illegitimate, asking for tallies, counterarguments, and factual accompaniments (B4). Others exhibited outcome-dependent trust: \textit{"when it passed, great; when it fails, I question legitimacy"} and called for a \textit{"trustworthy seal of approval"} to ground the process (B5). 
% Even those who trusted results wanted space to explain why their views diverged from the final outcome (B6).

% \textbf{Condition C:} Here, seeing the vote was the primary engine of legitimacy. One participant trusted because \textit{“I could see how people voted… usually you don’t get to see… you only know the outcome,” }adding that the visualization made it feel \textit{“legitimate… real… this isn’t fake”} (C2). Others accepted majority rule even when the decision didn't align with their vote—\textit{“it actually happens… all the time… it just doesn’t always work in your favor”} (C3). At the same time, they worried quick votes and sentence-long rationales on the voting form don’t reflect true considered views, and wanted ways to weigh pros/cons, and revisit their own choices (C4).

In \textbf{Condition C}, seeing the vote was the primary engine of legitimacy: \textit{"I could see how people voted…usually you don't get to see, you only know the outcome,"}. Others accepted majority rule even when it didn't favor them: \textit{"it actually happens all the time, it just doesn't always work in your favor"} (C3). Still, some worried that quick voting failed to capture considered views, and wanted ways to weigh pros and cons or revisit their own choices (C4).

% \textbf{Condition D} Trust was predictably the lowest in this case. One participant said they would be more trustful only if they understood \textit{“how the AI was programmed”} and its biases (D1). Another wanted percentages and reasons: \textit{“they not giving me nothing to go on… If it gave me more… with a percentage, I would have believed it more,”} adding they’d accept the decision as legitimate if they could \textit{“see… reasons”} and \textit{“could be persuaded”} (D2). Without these explanations, the process felt opaque and unconvincing.

In \textbf{Condition D}, trust was predictably lowest. D1 wanted to understand \textit{"how the AI was programmed"} and its biases; another wanted reasons: \textit{"they not giving me nothing to go on. If it gave me more with a percentage, I would have believed it more and could be persuaded"}(D2). The process thus felt opaque and unconvincing.

% Participants thus converged on three legitimacy requirements: (1) visibility of aggregation (who voted and how many—visualization, counts, percentages); (2) explanations (short, comprehensible rationales synthesizing why a proposal passed/failed, ideally grounded in participants’ own words); and (3) assurances of procedural integrity (vetted process, governance seal, and options to reform/revote). Where these element were missing, participants substituted community buy-in or majority-rule deference, but these were fragile and often dependent on whether the outcomes aligned with participant votes.

Participants thus converged on: (1) visibility on who voted and how; (2) explanations synthesizing why a proposal passed or failed, ideally in participants' own words; and (3) assurances on vetted process and options to revote.
% They otherwise deferred to majority-rule or community buy-in but this was contingent on whether outcomes aligned with their own votes.

\subsubsection{Conditional Willingness to Comply.} 
% Across participants, willingness to abide by outcomes depended less on agreement and more on whether decisions felt justified, transparent, and socially legitimate. Several rejected \textit{“bare majority”} rationales as insufficient: \textit{“The only explanation… is ‘because the majority of people supported this, thus we decided to pass.’ That’s not a reason… You didn’t say anything about how—if it’s good for business or not. (A2)”} Others said they would comply if the process had communal backing and provided justification. B1 noted they would \textit{“buy into it”} only if their community collectively adopted the system as their decision rule—\textit{“it’s more… me conforming to the social compact.”} Participants also framed compliance as civic or legal duty even when they disagreed: \textit{“I’m a rule follower… I might not agree with it, but… if you don’t have a choice…”} (A6). D3 emphasized honoring outcomes while exercising democratic recourse—“\textit{I have to honor and respect it… Can I protest it? Yes. Can I write letters…? All the time.”}

Across conditions, willingness to abide by outcomes depended less on agreement and more on whether decisions felt justified and transparent. Several rejected \textit{"bare majority"} rationales: \textit{"The only explanation is `because the majority of people supported this, thus we decided to pass.' That's not a reason…You didn't say anything about if it's good for business or not"} (A2). Others tied compliance to their community collectively backing the system.  Some framed it as civic duty: \textit{"I'm a rule follower…I might not agree with it, but if you don't have a choice"} (A6), while D3 emphasized honoring outcomes alongside recourse: \textit{"I have to honor and respect it. Can I protest it? Can I write letters? Yes."}

\subsubsection{Future Contexts for Using this Process.} 
% Participants distinguished clearly between contexts where they would welcome the tool and where they would resist it, largely depending on transparency, perceived stakes, and opportunities for dialogue. Many were comfortable with civic or community use, describing it as inclusive and low-friction. A7 underscored its promise for local decision-making that surfaces underrepresented voices without forcing public confrontation, while also warning that anonymity can backfire if people suspect the results are fabricated or controlled by unseen actors. A9 emphasized the tool’s potential to strengthen civic dialogue, noting that simply creating space to listen and understand can build a more compassionate community. Municipal elections served as a reference point for legitimacy—public counts, identifiable officials, and media scrutiny—leading participants to caution that withholding data or obscuring decision processes invites suspicion. Overall, they saw a strong fit for advisory or participatory civic use, where the tool amplifies diverse perspectives and supports deliberation, but only if paired with transparent evidence trails, credible oversight, and clear avenues to question and interpret outcomes.

Participants distinguished clearly between contexts where they would welcome the tool and where they would resist it, based on transparency, stakes, and opportunities for dialogue. Many were comfortable with civic or community use, seeing it as inclusive and low-friction. A7 highlighted its promise for local decision-making that surfaces underrepresented voices without forcing public confrontation, while warning that anonymity can backfire if results seem fabricated or controlled by unseen actors. A9 saw potential for building civic dialogue: simply creating space to listen can foster a more compassionate community. Municipal elections served as a legitimacy benchmark: public counts, identifiable officials, and media scrutiny led participants in conditions B and D to caution that obscuring decision processes invites suspicion. Participants thus saw strong fit for participatory civic use, but only with evidence trails, credible oversight, and avenues to question outcomes.

\subsection{Stances and Learning}

Participants described the process as a gateway to diverse perspectives, but how much those perspectives shifted opinions depended on whether they could both see others and trace their reasoning.

% In \textbf{Condition A}, exposure felt rich and human: participants said they were \textit{“hearing and reading [others’] opinion”} in ways that reframed disagreement as situational, not adversarial—\textit{“it was not really about somebody being right or wrong… it was about their own perspective from where they are in their life,”} which \textit{“made me really appreciate everybody’s opinion a lot more than I normally might”} (A4). For some, this broadened perspective but only modestly moved positions. A6 said hearing varied age groups and concerns \textit{“put things into perspective,”} especially around the cost of living implications of a \$30 minimum wage—\textit{“I think I was more… convinced… it should not be \$30 an hour… \$30 is a lot”}—even while still supporting raising the wage. Others described no shift but a clearer holistic understanding:  \textit{“I think it's a collectivistic sort of view of the situation, but it's, like… it's still individualized for me'' }(A7). Several wished they could re-vote after exposure: \textit{“I might change some of my answers… not dramatically”} (A9); A4 felt others \textit{“were starting to, like, even sway my opinion”} and wanted a chance to \textit{“talk about… whether you changed your stance.”}

In \textbf{Condition A}, hearing others reframed disagreement as situational rather than adversarial—\textit{"it wasn't really about somebody being right or wrong…it was about their own perspective from where they are in their life,"} which \textit{"made me really appreciate their opinion a lot more than I normally might"} (A4). A6 said hearing varied age groups \textit{"put things into perspective"} on the cost-of-living implications of a \$30 minimum wage. Others described no shift but a more holistic understanding: \textit{"it's a collectivistic sort of view, but it's still individualized for me"} (A7). Several wished to re-vote after exposure: \textit{"I might change some of my answers…not dramatically"} (A9), with A4 noting others were \textit{"starting to, like, even sway my opinion"}.

% In contrast, \textbf{Condition B} muted exposure and led to little movement. Many reported no change—\textit{“I’m set in my ways”} (B2); \textit{“everything was so opaque… once I was done… I kind of just forgot about it”}. The main reason was \textit{lack of access} to others’ views and arguments. B3 repeatedly asked to \textit{“hear what the other participants had said”} and for \textit{“more chances to talk,”} suggesting that shared nuance could show \textit{“we may all be feeling the same way… just [with] a little bit more nuance.”} B5 wanted the other side’s arguments to make sense of disagreement, and B6 emphasized these topics aren’t \textit{“black and white,”} arguing for dialog-then-vote. Without those affordances, exposure rarely translated into opinion shift.

In \textbf{Condition B}, limited exposure muted learning and led to little movement: \textit{"everything was so opaque…once I was done…I kind of just forgot about it."}  The core barrier was lack of access to others' views: B3 repeatedly asked to \textit{"hear what the other participants had said"} since sharing could reveal \textit{"we may all be feeling the same way…just [with] a little bit more nuance."}  B5 wanted opposing arguments to make sense of disagreement; B6 argued these topics aren't \textit{"black and white"} and called for dialog before voting. 
% Without these affordances, exposure rarely translated into opinion shift.

% \textbf{Condition C} again created fertile exposure—and here we saw both \textit{shifts} and \textit{reinforcement}. Some explicitly changed their minds after hearing work experiences specific to certain sectors. Others said the process \textit{reinforced} prior views by clarifying both \textit{“flaws”} and \textit{“strengths”} (C4), or by encouraging balanced consideration without ultimately shifting views. C4 also offered a critical reading of what she perceived as \textit{pessimistic} stances, noting some \textit{“project[ed]… their disdain for the things that they went through,”} when forming an opinion rather than thinking reasonably. Several emphasized that hearing actual voices and seeing multiple sides prompted deeper reflection even when it didn’t change a vote—C3 described the process a personal \textit{“mental project,”} iteratively weighing both sides before concluding.

\textbf{Condition C} again created fertile exposure, producing both shifts and reinforcement. Some explicitly changed their minds after hearing sector-specific work experiences; others found the process reinforced prior views by clarifying both \textit{"flaws"} and \textit{"strengths"} (C4) without ultimately shifting them. C4 also offered a critical reading of what she saw as pessimistic stances, noting some \textit{"project[ed] their disdain for the things that they went through"} rather than reasoning from broader grounds. Several emphasized that hearing actual voices and seeing multiple perspectives prompted deeper reflection on weighing both sides even without changing a vote.

% Finally, in \textbf{Condition D}, exposure was weakest but still nudged \textit{meta-level openness}. D1 said seeing their opinion go against the majority \textit{“made me question my stance,”} and the process would \textit{“help with the acceptance portion.”} D2 said they \textit{needed} to \textit{“hear… people and their experience”} and wanted simple aggregates and reasons (\textit{“10 people… 8 felt this way… why?”}) to understand divergence. 

In \textbf{Condition D}, exposure was weakest but still nudged meta-level openness. D1 noted that seeing their opinion go against the majority \textit{"made me question my stance"} and would \textit{"help with the acceptance portion."}  D2 felt they \textit{needed} to \textit{"hear people and their experience,"} wanting numbers and reasons to explain divergence.

Opinion shift was thus most likely when participants could hear lived context as opposed to obscuring others' thinking.

\subsection{Role of AI in the Process}
\label{results:AI-interviewer}
% Participants saw clear value in AI but emphasized oversight, transparency, and context. Many enjoyed speaking with the AI interviewer—finding it smooth, motivating, or easier than talking to a human who might judge them (A4, B2); some even treated it \textit{“like a person”} (A6). Comfort was conditional: a few avoided sensitive topics if identifiable (B3) or missed emotional nuance (A8). For the visualization, participants endorsed AI for scalability and aggregation (A1) and some viewed it as potentially fairer than humans if trained on diverse input (D2). Core concerns focused on opacity and fidelity: several wanted to know how the system was built and tested for bias (D1), worried about “black box” summaries that distort intent (C4) or lean ideologically (B5), and feared AI might \textit{“change my words”} (A5), preferring to verify against transcripts (D3). Overall, they argued for AI as support for collection and synthesis—not as final decision-maker—especially in high-stakes contexts.

Participants saw clear value in AI but emphasized oversight and transparency. Many found it easier than a human who might judge them (A4, B2), but still \textit{"like a person"} (A6). Some avoided sensitive topics over anonymity concerns (B3) or missed emotional nuance (A8). For the visualization, participants endorsed AI for scalability and aggregation (A1), viewing it as potentially fairer than humans if trained on diverse input (D2). Core concerns centered on opacity and fidelity: participants wanted to know how the system was built and tested for bias (D1), worried about \textit{"black box"} summaries distorting intent (C4) or leaning ideologically (B5), and feared AI might \textit{"change my words"} (A5), preferring verification against transcripts (D3). Participants thus positioned AI for collection and synthesis, not as final decision-maker especially in high-stakes contexts.

\section{Discussion}
Our goal was to design a system strengthening process conditions to help participants feel heard, respected, and willing to accept decisions even when they go against them. We examine how AI-powered tools can enhance trust, understanding, and recognition across diverse experiences.
We built on best practices from HCI \cite{considerit, PolicyScape, reflection_nudge, yeo2024helpmereflect}, social psychology \cite{kubin, kessler2023hearing}, and deliberation \cite{oxford_delib, gastil2005deliberative, setala2018mini}, centering authentic experiences while achieving scale via semi-structured AI interviews and LLM extraction of relevant experiences. We discuss our tool's impact on losers' consent objectives and identify areas for improvement and future work.

\subsection{Impact on Process Trust and Legitimacy}
Previous work identified trust as a challenge in public deliberation platforms, with \citet{considerit} noting participants ``wanted to know more about the people who were adding the points.'' Process trust and legitimacy are crucial for accepting disagreeable outcomes. 
Our visualization addressed this: participants valued hearing backgrounds and seeing the support spectrum. However, qualitative feedback revealed limits: procedural options like ability to repeal/reform decisions were lacking. Future work could blend our approach with tools for iterative consensus refinement \cite{habermas}.

Both the AI interview and visualization increased whether participants felt heard. However, some doubted their avatars would actually be viewed by others, limiting representation impact. Features like showing profile views could address this but risk encouraging performative input, like polarizing rhetoric attracting engagement on social media \cite{outgroup_polarizing}. Balancing authentic expression with genuine recognition is an important direction for future research.

\subsection{Impact on Social Cohesion}
Participants using the visualization reported higher social cohesion on most measures. To support this, we incorporated reflective prompts after hearing others' input \cite{reflection_nudge, yeo2024helpmereflect}. Unlike prior text-based online deliberation systems \cite{considerit, opinion_space, PolicyScape, small2021polis}, we included real voice audio. Participants valued hearing tone and cadence, which improved perceptions of others. Amid concerns about LLM-fabricated text, unedited audio ensured authenticity and provenance. This addresses \citet{considerit}'s finding that participants want more context: hearing backgrounds before policy views helped grasp underlying reasoning, addressing deliberative systems' common tendency to strip contextual richness \cite{PolicyScape}.
Research also shows participants in online deliberations object to content deemed false, misleading, or insufficiently persuasive \cite{considerit}. While our visualization had strong positive effects on respect and perceived rationality, reception varied. Content connecting personal experiences to policy implications or engaging counter-arguments was well-received; one-off anecdotes were sometimes seen as self-centered. This raises questions for better connecting personal experiences effective at bridging divides \cite{kubin} to concrete policy outcomes.

While most social cohesion measures improved, participants did \textit{not} report significantly greater willingness to engage with others. Though hearing personal experiences encouraged this for some, others felt uncertain if further interaction would succeed with very different perspectives. We saw no full "backfire" effect, but recognizing diversity didn't universally lead to more positive impressions.

\subsection{Impact on Stances and Learning}
Learning from others' perspectives and updating opinions based on this information is central to productive deliberation \cite{oxford_delib, gastil2005deliberative}. We adopted HCI approaches like mapping agreement spectrums \cite{opinion_space} and reflection nudges for perspective-taking \cite{yeo2024helpmereflect, yeo2025enhancing}. Participants viewing our visualization reported learning about the topic and small self-reported stance shifts, though pre/post measurements showed no actual policy support changes. Instead, participants \textit{understood} others' perspectives, broadening their views without meaningful stance shifts. Some clarified and reflected on their own stances, even when unchanged. We believe prioritizing openness to different perspectives \cite{Farag2022Opening} over persuasion is important for future deliberative technologies, especially on controversial issues.

\section{Limitations}
\subsection{Study Limitations}
While our \(2 \times 2\) design examined the AI interviewer and visualization's impact, we couldn't isolate individual visualization features such as support distribution displays, types of background information, presentation of experiences vs. beliefs, or audio vs. text modalities. Our learning measure relied on self-reports, capturing perceived understanding but potentially missing objective learning from seeing others' viewpoints. Future work could use more direct measures like search-as-learning methods~\cite{vakkari2016searching}.
The complexity and cost of data collection limited sample size and statistical power. Our study only examined ``losing'' scenarios, when outcomes didn't align with preferences, so findings don't generalize to participants whose preferred outcome prevailed. Conducting ablation studies across all these factors was infeasible. Future studies can examine how randomizing outcomes influences trust, legitimacy, and social connection. Our aim was to demonstrate how AI can implement established HCI best practices in a functional, scalable system.

\subsection{Platform Limitations}
Participants appreciated the AI interviewer's affordances (\ref{results:AI-interviewer}) but acknowledged privacy risks. AI interviews can carry social and perceptual biases from voice, timing, and embodiment \cite{biswas2024hi}; we used the same female voice as prior work \cite{park2024generativeagentsimulations1000} but didn't study gender/accent effects on interactions. Our study focused on inferring interview content rather than human-agent interaction dynamics. Future work should examine implications of deploying AI interviewers with these characteristics in decision-making settings.

Our visualization drew on deliberative democracy practices: exposing and encouraging holistic understanding of diverse perspectives and hearing others' views in their own words. However, back-and-forth dialogue and expert learning phases were absent. Given privacy risks with real voice recordings, future iterations could integrate watermarking, provenance indicators, or prosody-preserving anonymization. Though reducing support and experience to single dimensions aids interpretability, it simplifies inherent nuance in each, underscoring the need for richer encodings in the future.

While our platform cannot replace fully deliberative processes, it demonstrates how AI can enhance trust in collective decisions, process legitimacy, and social connectedness—fundamental building blocks of healthy democracy \cite{levi2000political, why_trust_matters, why_people_obey_law}. With trust in institutions and other citizens near all-time lows \cite{silver2025trust, pew2024mistrust}, technology strengthening these connections is crucial. This work could inspire future research using technology to enhance collective decision-making.

% While these are important, an under-explored direction of research is how we can use AI to also enhance trust in collective decisions and process legitimacy, as well as social connectedness of participants themselves, all of which are fundamental building blocks of a healthy democracy \cite{levi2000political, why_trust_matters, why_people_obey_law}. At a time when trust in institutions and other citizens is near all time lows \cite{silver2025trust, pew2024mistrust}, focusing on technology that strengthens these connections is even more important. We hope this work inspires future research that aims to use technology to enhance these other goals of collective decision-making systems.

% \begin{acks}
% \end{acks}

\bibliographystyle{ACM-Reference-Format}
\bibliography{sample-base}
% \section{Declaration of Generative AI Software Tools in the Writing Process}

% During the preparation of this work, the author(s) used OpenAI's GPT models for copy-editing portions of authors' text to improve clarity and brevity. After using this tool, the authors reviewed and edited the content as needed and takes full responsibility for the content of the publication.

%TC:ig
\appendix
\section{Predicted Support Validation}
\label{app:predicted-support}

We computed the accuracy, correlation, and MAE of the LLM predicted support and what participants voted prior to interacting with the interface and making a final decision. To compute the accuracy, we simply binarized both the initial vote and predicted support. Above fifty on predicted support we considered to be positive, and below was negative. Since the stance prior to voting was on a six-point scale, we binarized this by considering votes four and higher to positive and three and lower as negative. We also computed the MAE by regression the predicted support (which is on a scale of 0-100) on the prior stance to properly scale the prior stance. We then compute the MAE on the scaled prior stance.
\begin{table}[htbp]
  \centering
  \begin{tabular}{lccc}
    \toprule
    Recommendation ID & Accuracy & Correlation & MAE \\
    \midrule
    74      & 0.81 & 0.79 & 17 \\
    75      & 0.81 & 0.61 & 20 \\
    76      & 0.84 & 0.77 & 16 \\
    \midrule
    Average & 0.82 & 0.72 & 17 \\
    \bottomrule
  \end{tabular}
  \caption{Accuracy, correlation, and mean absolute error (MAE) by recommendation.}
  \label{tab:rec_performance}
\end{table}

\section{Prompts for Visualization}
\label{app:viz-prompts}

\subsection{Policy Related Prompts}
\subsubsection{\textbf{Prediction Prompts}}
\label{app:pred_prompt}
\begin{lstlisting}[style=promptstyle]
You are an assistant that analyzes participant transcripts to predict 
how much a participant would agree with a specific recommendation, 
based on their prior statements and experiences.

Your task is to return a JSON object that includes:
1. A brief explanation (max 100 words) of why the participant would 
   agree or disagree with the recommendation.
2. A predicted agreement level (0-100), where 0 means total disagreement 
   and 100 means complete agreement.
3. A confidence score (0-100) reflecting how confident you are in your 
   prediction.

Here is the transcript for participant {display_name}:
{transcript}

Here is the recommendation:
"{rec_text}"

Return only a valid JSON object in the following format:
{
    "reasoning": "Your explanation here (max 100 words)",
    "predicted_agreement": <integer between 0 and 100>,
    "confidence_score": <integer between 0 and 100>
}
\end{lstlisting}
\subsubsection{\textbf{Selecting Utterances Prompt}}
\label{app:select_utt}
\begin{lstlisting}[style=promptstyle]
You are an assistant that identifies the best supporting evidence from 
a participant's transcript that relates to a specific recommendation.

Given the transcript, reasoning for predicted agreement, and the 
recommendation, identify the TOP 2 most relevant utterances that best 
support the prediction.
Bias towards personal experiences that are relevant to the 
recommendation. Make sure at least one of the utterance is a personal 
experience related to the recommendation. 
Try not to select opinion statements (e.g. I think, I feel, I believe, 
etc.). Rather, select statements that reflect a person's life 
experiences, how they would feel about the recommendation,
or how it would impact them or their family/community.

For each identified utterance, provide:
- The utterance ID
- The utterance text with key phrases highlighted using <b> tags
- A brief explanation of why this utterance is relevant

Here is the transcript:
{transcript}

Here is the reasoning for predicted agreement:
{reasoning}

Here is the recommendation:
"{rec_text}"

Return only a valid JSON object in the following format:
{
    "evidence": [
        {
            "utterance_id": <integer>,
            "utterance_text_bolded": "The utterance text with highlighted key phrases",
            "relevance_explanation": "Brief explanation of why this evidence is relevant"
        },
        {
            "utterance_id": <integer>,
            "utterance_text_bolded": "The utterance text with highlighted key phrases", 
            "relevance_explanation": "Brief explanation of why this evidence is relevant"
        }
    ]
}

Important guidelines:
- Select exactly 2 utterances (no more, no less)
- Bias towards personal experiences, stories, or situations that are 
  directly relevant to the recommendation
- Be selective - only include the most relevant and compelling evidence
- Make sure all utterance IDs exist in the transcript
- Highlight key phrases using <b> tags to emphasize the most relevant parts
- Provide brief explanations for why each utterance is relevant
\end{lstlisting}
\subsubsection{\textbf{Scoring Utterances Prompt}}
\label{app:score_utt}
\begin{lstlisting}[style=promptstyle]
You are a senior academic researcher who reviews qualitative data for 
rigor and quality. You have extensive experience evaluating how relevant 
and deep a participant's experiences are to a specific recommendation.

Analyze the provided experiences and rate them on:
1. Opinion vs. Experience (0-100): Is this a personal story or 
   experience, or is the person sharing how they or their community 
   would be impacted? If so, return a high score to indicate that this 
   person is sharing experiences. If they are mostly stating their 
   beliefs (e.g. "I think" "I believe"), then report a low score to 
   indicate these are opinions.
2. Relevance (0-100): How directly related are these experiences to the 
   recommendation? This is not about the opinion expressed; rather, is 
   this person sharing stories or explaining how they or their community 
   would be impacted?

- 90-100: Direct, first-hand experience that directly impacts the 
          person's stance on this exact recommendation
- 70-89: Related experience with clear, obvious connection to the 
         recommendation
- 50-69: Somewhat related experience but requires reasoning to connect 
         to the recommendation
- 30-49: Tangentially related, minimal impact on stance
- 0-29: Unrelated or generic statements

3. Depth (0-100): How detailed, specific, and meaningful are these 
   experiences?

- 90-100: Specific details, concrete examples, clear timeline, emotional 
          impact described
- 70-89: Good detail with some specifics, clear narrative
- 50-69: Some detail but lacks specificity or emotional depth
- 30-49: Vague or surface-level description
- 0-29: Generic statements with no real detail

Consider:
- Relevance: Do the experiences directly relate to the recommendation's 
  topic? Would they likely influence the person's stance?
- Depth: Are the experiences specific, detailed, and meaningful? Do 
  they show real understanding or just surface-level mentions?

Here are the experiences:
{experiences}

Here is the recommendation:
"{rec_text}"

IMPORTANT CALIBRATION INSTRUCTIONS:
Be conservative in your scoring. Only award high scores (80+) to 
experiences that are:
1. Directly related to the recommendation's specific topic
2. Include specific details, dates, locations, or concrete examples
3. Show clear personal impact or emotional connection
4. Would genuinely influence someone's stance on this recommendation? 
   Even if they have a polar opposite view?

Default to lower scores when in doubt. It's better to underestimate than 
overestimate quality.

Return only a valid JSON object in the following format:
{
    "opinion_vs_experiences": <integer between 0 and 100>,
    "relevance_score": <integer between 0 and 100>,
    "depth_score": <integer between 0 and 100>,
    "explanation": "Brief explanation of your scoring for each metric (max 50 words)"
}
\end{lstlisting}
\subsubsection{\textbf{Utterance Summary Prompt}}
\label{app:utt_summary}
\begin{lstlisting}[style=promptstyle]
You are an assistant that creates concise summaries of participant 
experiences.

Summarize the provided experiences in a clear, informative way that 
captures the key points and their significance. Focus on the most 
important aspects that would be relevant to understanding the 
participant's perspective.

Here are the experiences:
{experiences}

Return only a valid JSON object in the following format:
{
    "summary": "Your summary here (max 100 words)"
}

Guidelines:
- Be concise but comprehensive
- Focus on the most relevant and significant experiences
- Maintain the participant's voice and perspective
- Highlight experiences that would influence their views on related topics
\end{lstlisting}
\subsection{Life Prompts}
\label{app:life_prompts}
\subsubsection{\textbf{Life Utterance Prompt}}
\begin{lstlisting}[style=promptstyle]
You are a helpful assistant that selects the most relevant utterances 
from the transcript that would help someone understand the participant's 
life. 

Select at most two utterances from the transcript. These should be 
excerpts that serve as an introduction to who this person is and what 
they are like. 

IMPORTANT: Make sure you include the utterance id in the JSON. Include 
only a single utterance id (an integer) for each part of the narrative.

Here is the transcript of the participant {display_name}:
{transcript}

Return the narrative format in a list of JSON with top level key 
"utterances". Each json should have the following format:
{
    "utterances": [
        {   
            "utterance_id": "The utterance id for the first part.",
            "interviewee_utterance": "The utterance text for the first part"
        },
        {   
            "utterance_id": "The utterance id for the second part.",
            "interviewee_utterance": "The utterance text for the second part"
        }
    ]
}
The list should flow naturally. Remember, it should be a list of JSONs 
of at most two. 

Use the participant's name in the narrative. Remember to pick 
sufficiently long excerpts from the transcript for the 
interviewee_utterance.
\end{lstlisting}
\subsubsection{\textbf{Life Summary Prompt}}
\begin{lstlisting}[style=promptstyle]
You are a helpful assistant that generates a summary of the following 
statements from a person. 

The statements are from the person about their life. Please return a 
short, informative summary of who this person is. Describe the person's 
life in a way that is easy to understand and engaging, but stay very 
close to what the person said. 

The person's name is {display_name}, which you should use in the summary. 
Keep the summary under 50 words.

Here are the utterances:
{utterances}

Return the summary in a list of JSON with top level key "summary". Each 
json should have the following format:
{
    "summary": "The summary of the utterances"
}

Remember to use the person's name in the summary.
\end{lstlisting}

\section{Sampling Profiles}
\label{app:samp_prof}
In our experimental setup, we wanted to particularly focus on how trust and social cohesion were affected when the collective decision went against what the participant desired. Thus, in the treatment condition we needed to show a set of profiles that aligned with the decision was shown to the participant. 

To do this, we used a constrained optimization to assign a set of weights to the different profiles depending on their level of support and the target mean support. In the case where a participant was \textit{against} the proposal, we set the target mean support at 75. If they voted \textit{for} the proposal, the target mean was set at 25. We chose to select profiles rather than change the predicted scores of profiles because we thought that inconsistencies between what people said and their predicted support would decrease trust in the system (i.e., if a person expressed a strong desire to raise the minimum wage and their predicted support was medium or low, this would decrease trust in the system's accuracy). Thus the constrained optimization was set as follows. Given:
\begin{itemize}
    \item A set of support values: 
    \[
    s = [s_1, s_2, \dots, s_n]
    \]
    \item A target mean: 
    \[
    \mu_{\text{target}}
    \]
    \item Optimization variables: the weights
    \[
    w = [w_1, w_2, \dots, w_n]
    \]
\end{itemize}

\noindent
\textbf{Objective Function:}  
A trivial solution to this problem is to assign all of the weight to a value equal to the target mean. Thus, we want the weights to be as close as possible to uniform weights 
\(w_{\text{uniform}} = \frac{1}{n}\) and set the objective function as such. Thus, we minimize the squared L2 distance:
\[
\boxed{
\min_{w} \; \sum_{i=1}^n \left( w_i - \frac{1}{n} \right)^2
}
\]

\noindent
\textbf{Constraints:} We set the constraints such that the weights should sum to one, range between zero and one, and when applying the weights to the scores, it yields the target mean.
\begin{align}
    \sum_{i=1}^n w_i &= 1 && \text{(weights sum to one)} \\
    \sum_{i=1}^n w_i \, s_i &= \mu_{\text{target}} && \text{(target mean constraint)} \\
    0 \leq w_i &\leq 1, \quad \forall i && \text{(bounds on weights)}
\end{align}

\noindent
\textbf{Final Optimization Problem:}
\[
\begin{aligned}
\min_{w \in \mathbb{R}^n} \quad & \sum_{i=1}^n \left( w_i - \frac{1}{n} \right)^2 \\
\text{subject to} \quad 
& \sum_{i=1}^n w_i = 1, \\
& \sum_{i=1}^n w_i \, s_i = \mu_{\text{target}}, \\
& 0 \leq w_i \leq 1 \quad \forall i.
\end{aligned}
\]
We then used these weights to take a biased sample of profiles. Since the original distribution of support for each issue was often skewed, we could not sample of set of profiles in each case that exactly lead to either a mean support of 75 or 25. However, there was enough diversity to sample a set of profiles that aligned that aligned with the decision (i.e. greater than 50 for pass and less than 50 for not pass). Below we should the mean support of the sampled profiles depending on whether we displayed the decision as passing or not passing for each proposal.

% \begin{table}[h!]
% \centering
% \small
% % \renewcommand{\arraystretch}{1.3}
% \begin{tabular} 
% % \begin{tabular}{\textwidth}{@{}Xcc@{}} 
% % \toprule
% \textbf{Proposal} & \textbf{Against Mean} & \textbf{Pro Mean} \\ 
% \midrule
% The federal minimum wage should be raised to \$30 an hour. & 41.1 & 63.3 \\
% Companies should strongly prioritize hiring domestically before considering foreign applicants. & 44.0 & 67.8 \\
% Race and gender should be used in hiring decisions to combat inequality in the workplace. & 31.8 & 54.7 \\
% % \bottomrule
% \end{tabular}
% \caption{Pro and against means for sampled profiles.}
% \end{table}

\begin{table}[h!]
\centering
\small
\begin{tabularx}{\columnwidth}{@{}Xcc@{}}
\toprule
\textbf{Proposal} & \textbf{Against Mean} & \textbf{Pro Mean} \\
\midrule
The federal minimum wage should be raised to \$30 an hour. & 41.1 & 63.3 \\
Companies should strongly prioritize hiring domestically before considering foreign applicants. & 44.0 & 67.8 \\
Race and gender should be used in hiring decisions to combat inequality in the workplace. & 31.8 & 54.7 \\
\bottomrule
\end{tabularx}
\caption{Pro and against means for sampled profiles.}
\end{table}

\section{Number of Potential Featured Profiles}
\label{app:featured_profiles_bucket}

We show the number of potential featured profiles for each recommendation in each bucket of support (low, medium, and high support) in \autoref{tab:policy_responses}. When interacting with the visualization, a user is shown one randomly sampled profile from each bucket for each recommendation. We also considered other approaches, such as using percentiles to identify segments of the distribution to sample from. However, the initial distribution of opinions for the topics was still biased either for or against each recommendation, meaning that percentile-based sampling would have led participants to see more profiles with a particular slant. Thus, using equal-sized buckets offered the simplest and most reliable way to ensure that participants saw a sufficiently diverse range of opinions.

\begin{table}[ht]
\centering
\small
\begin{tabularx}{\columnwidth}{@{}Xccc@{}}
\toprule
\textbf{Statement} & \textbf{Low (0--33)} & \textbf{Mid (33--66)} & \textbf{High (66+)} \\
\midrule
The federal minimum wage should be raised to \$30 an hour. & 5  & 10 & 23 \\
Companies should strongly prioritize hiring domestically before considering foreign applicants. & 7  & 4  & 8  \\
Race and gender should be used in hiring decisions to combat inequality in the workplace. & 17 & 9  & 8  \\
\bottomrule
\end{tabularx}
\caption{Responses categorized by support levels for different policy statements.}
\label{tab:policy_responses}
\end{table}

\section{Item Reliability within Concept}
\label{app:survey_reliability}

In \autoref{tab:cronbach} we report Cronbach's alpha for each of our concepts. Most concepts had high item consistency with the exception of  the ``Felt Heard'' concept. The statement ``My own opinions and experiences were not reflected in any way by the proposals.'' in particular a low correlation with both ``I felt that my input was valuable in this process.'' and ``I felt heard during this process.'' While we are not sure what caused this low correlation, we suspect this question was poorly worded and thus may have been difficult for participants to answer.
\begin{table}[ht]
\centering
\small
\begin{tabularx}{\columnwidth}{@{}lXcc@{}}
\toprule
\textbf{Domain} & \textbf{Concept} & \textbf{Cronbach's $\alpha$} & \textbf{N items} \\
\midrule

\multirow{4}{*}{Process Legitimacy} 
  & Adherence               & 0.93 & 5 \\
  & Understood Decisions    & 0.88 & 4 \\
  & Trust                   & 0.78 & 3 \\
  & Felt Heard              & 0.55 & 3 \\

\addlinespace

\multirow{6}{*}{Social Cohesion}
  & Learn about others      & 0.93 & 4 \\
  & Respect                 & 0.87 & 5 \\
  & Rational                & 0.86 & 3 \\
  & Respect Pluralism       & 0.83 & 5 \\
  & Commonalities           & 0.75 & 4 \\
  & Perspective-taking      & 0.66 & 3 \\

\addlinespace

\multirow{3}{*}{Stances and Learning}
  & Learning about topics         & 0.98 & 9 \\
  & Change stance (self report)   & 0.82 & 3 \\
  & Stance certainty              & 0.79 & 9 \\

\bottomrule
\end{tabularx}
\caption{Cronbach's $\alpha$ for concepts grouped by domain.}
\label{tab:cronbach}
\end{table}

\onecolumn

\section{Regression Results}
\label{app:reg_results}

Here we present the full set of regression results across all concepts and items. For concepts, we report the p-values. For items, since we are testing multiple items per concept, we apply the Benjamini-Hochberg correction and report the corresponding q-value. 
\subsection{Process Legitimacy}
\begin{table}[H]
\centering
\footnotesize
\begin{tabular}{p{2cm}p{3cm}cccccccccc}
\toprule
\multicolumn{2}{c}{} & \multicolumn{3}{c}{\textbf{Visualization}} & \multicolumn{3}{c}{\textbf{Interview}} & \multicolumn{3}{c}{\textbf{Interaction}} & \\
\cmidrule(lr){3-5} \cmidrule(lr){6-8} \cmidrule(lr){9-11}
\textbf{Category} & \textbf{Outcome} & Coef. & 90\% CI & $p$ & Coef. & 90\% CI & $p$ & Coef. & 90\% CI & $p$ & $R^2_{adj}$ \\
\midrule
\multirow{4}{*}{Process Legitimacy} 
& Adherence & 0.34* & [0.04, 0.64] & 0.06 & 0.11 & [-0.20, 0.42] & 0.55 & -0.23 & [-0.66, 0.19] & 0.36 & 0.24 \\
& Felt Heard & 0.29* & [0.04, 0.54] & 0.06 & 0.51*** & [0.25, 0.76] & 0.001 & -0.20 & [-0.56, 0.15] & 0.34 & 0.21 \\
& Trust & 0.71*** & [0.41, 1.01] & <.001 & 0.19 & [-0.11, 0.50] & 0.30 & -0.53** & [-0.95, -0.11] & 0.04 & 0.13 \\
& Understood Decisions & 0.77*** & [0.47, 1.07] & <.001 & 0.19 & [-0.12, 0.50] & 0.31 & -0.32 & [-0.74, 0.11] & 0.22 & 0.18 \\
\bottomrule
\end{tabular}
\caption{Concept outcomes associated with adherence to participatory processes across visualization and interview modes.}
\end{table}
\begin{table}[H]
\centering
\footnotesize
\begin{tabular}{p{2cm}p{3.5cm}ccccccccc}
\toprule
\multicolumn{2}{c}{} & \multicolumn{3}{c}{\textbf{Visualization}} & \multicolumn{3}{c}{\textbf{Interview}} & \multicolumn{3}{c}{\textbf{Interaction}} \\
\cmidrule(lr){3-5} \cmidrule(lr){6-8} \cmidrule(lr){9-11}
\textbf{Category} & \textbf{Outcome} & Coef. & 90\% CI & $q$ & Coef. & 90\% CI & $q$ & Coef. & 90\% CI & $q$ \\
\midrule
\multirow{5}{*}{Adherence} 
& I would be willing to abide by the specific proposal decisions generated from this process. & 0.88 & [0.26, 1.49] & 0.10 & 0.43 & [-0.20, 1.06] & 0.81 & -0.84 & [-1.73, 0.03] & 0.50 \\
& Generally, I would be willing to abide by proposal decisions from this process (on other issues). & 0.64 & [0.01, 1.26] & 0.10 & 0.38 & [-0.26, 1.03] & 0.81 & -0.69 & [-1.57, 0.16] & 0.83 \\
& Even if I disagreed with certain proposal decisions, I would view them as being legitimately created. & 0.69 & [0.08, 1.29] & 0.12 & 0.04 & [-0.55, 0.63] & 0.93 & -0.11 & [-0.96, 0.75] & 0.83 \\
& Even if I disagreed with certain proposal decisions, I would still respect them. & 0.60 & [0.03, 1.17] & 0.12 & 0.03 & [-0.55, 0.62] & 0.93 & -0.22 & [-1.03, 0.59] & 0.83 \\
& Even if I disagreed with certain proposal decisions, I would still adhere to them. & 0.14 & [-0.44, 0.72] & 0.70 & 0.10 & [-0.50, 0.70] & 0.93 & -0.19 & [-1.01, 0.64] & 0.83 \\
\multirow{3}{*}{Felt Heard} 
& I felt that my input was valuable in this process. & 0.98** & [0.39, 1.57] & 0.02 & 0.63 & [0.03, 1.23] & 0.13 & -1.13* & [-2.07, -0.29] & 0.08 \\
& My own opinions and experiences were not reflected in any way by the proposals. & 0.42 & [-0.27, 1.11] & 0.48 & -0.04 & [-0.75, 0.67] & 0.93 & -0.09 & [-0.88, 0.89] & 0.88 \\
& I felt heard during this process. & 0.11 & [-0.36, 0.59] & 0.70 & 1.99*** & [1.50, 2.48] & <.001 & 0.14 & [-0.54, 0.82] & 0.88 \\
\multirow{3}{*}{Trust \& Legitimacy} 
& I trust the legitimacy of the votes that came of this process. & 1.47*** & [0.88, 2.07] & <.001 & 0.58 & [-0.03, 1.18] & 0.36 & -1.14* & [-1.98, -0.30] & 0.08 \\
& I think the votes on these proposals were well-informed. & 1.24*** & [0.60, 1.87] & 0.003 & 0.37 & [-0.36, 1.02] & 0.53 & -1.03* & [-1.89, 0.06] & 0.09 \\
& I would be hesitant to rely on this process to make and vote on proposals on other issues. & 0.93** & [0.26, 1.60] & 0.02 & 0.03 & [-0.66, 0.71] & 0.95 & -0.54 & [-1.49, 0.40] & 0.34 \\
\multirow{4}{*}{Understood Decisions} 
& I understood how the proposal decisions were made. & 1.11*** & [0.49, 1.73] & 0.003 & -0.38 & [-1.02, 0.25] & 0.42 & -0.69 & [-1.57, 0.19] & 0.40 \\
& I had a strong understanding of how participants' opinions and experiences informed the decisions on the proposals. & 1.55*** & [0.93, 2.16] & <.001 & 0.61 & [-0.02, 1.24] & 0.42 & -0.44 & [-1.31, 0.42] & 0.40 \\
& I was unsure how participants' opinions and experiences aligned with or diverged from each proposal. & 1.55*** & [0.87, 2.23] & <.001 & -0.11 & [-0.81, 0.59] & 0.79 & -0.54 & [-1.50, 0.42] & 0.40 \\
& I understood why each proposal decision was made. & 1.24*** & [0.64, 1.83] & 0.001 & 0.41 & [-0.20, 1.02] & 0.42 & -0.54 & [-1.38, 0.31] & 0.40 \\
\bottomrule
\end{tabular}
\caption{Item outcomes associated with adherence to participatory processes across visualization and interview modes.}
\end{table}
\subsection{Social Cohesion}
\begin{table*}
\centering
\footnotesize
\setlength{\tabcolsep}{4pt}
\begin{tabular}{p{2.2cm}p{3.2cm}cccccccccc}
\toprule
\multicolumn{2}{c}{} 
& \multicolumn{3}{c}{\textbf{Visualization}} 
& \multicolumn{3}{c}{\textbf{Interview}} 
& \multicolumn{3}{c}{\textbf{Interaction}} 
& \\
\cmidrule(lr){3-5} \cmidrule(lr){6-8} \cmidrule(lr){9-11}
\textbf{Category} & \textbf{Outcome} 
& Coef. & 90\% CI & $p$ 
& Coef. & 90\% CI & $p$ 
& Coef. & 90\% CI & $p$ 
& $R^2_{adj}$ \\
\midrule
\multirow{10}{*}{Social Cohesion} 
& Commonalities & 0.40** & [0.16, 0.64] & 0.01 & 0.04 & [-0.21, 0.29] & 0.79 & -0.03 & [-0.37, 0.31] & 0.87 & 0.23 \\
& Inclusion of Self in Others & 0.48** & [0.12, 0.84] & 0.03 & 0.11 & [-0.26, 0.48] & 0.63 & -0.18 & [-0.70, 0.33] & 0.56 & 0.11 \\
& Feel Connected & 0.82*** & [0.48, 1.16] & $<$.001 & 0.18 & [-0.17, 0.53] & 0.40 & -0.50* & [-0.98, -0.02] & 0.09 & 0.22 \\
& Willing to Interact & -0.14 & [-0.49, 0.20] & 0.49 & 0.05 & [-0.30, 0.40] & 0.81 & 0.14 & [-0.35, 0.62] & 0.64 & 0.21 \\
& Curious About Others & -0.10 & [-0.46, 0.27] & 0.65 & -0.32 & [-0.70, 0.05] & 0.16 & 0.06 & [-0.46, 0.58] & 0.85 & 0.10 \\
& Learn About Others & 1.42*** & [1.16, 1.67] & $<$.001 & 0.29* & [0.02, 0.55] & 0.07 & -0.35 & [-0.72, 0.01] & 0.11 & 0.47 \\
& Perspective-taking & 0.44** & [0.18, 0.70] & 0.01 & 0.03 & [-0.24, 0.30] & 0.87 & 0.09 & [-0.29, 0.46] & 0.70 & 0.24 \\
& Rational & 0.58*** & [0.29, 0.88] & $<$.001 & -0.27 & [-0.58, 0.03] & 0.14 & -0.07 & [-0.49, 0.35] & 0.78 & 0.26 \\
& Respect & 0.58*** & [0.34, 0.83] & $<$.001 & 0.02 & [-0.22, 0.27] & 0.87 & -0.35* & [-0.70, -0.01] & 0.09 & 0.41 \\
& Respect Pluralism & 0.29** & [0.10, 0.49] & 0.01 & 0.03 & [-0.16, 0.23] & 0.77 & -0.25 & [-0.52, 0.03] & 0.14 & 0.58 \\
\bottomrule
\end{tabular}
\caption{Concept outcomes associated with social cohesion across visualization and interview modes.}
\end{table*}
{\footnotesize{
\begin{longtable}{p{1.8cm}p{3.5cm}ccccccccc}
\caption{Outcomes associated with perceptions of social cohesion across visualization and interview modes.} \\
\toprule
\multicolumn{2}{c}{} & \multicolumn{3}{c}{\textbf{Visualization}} & \multicolumn{3}{c}{\textbf{Interview}} & \multicolumn{3}{c}{\textbf{Interaction}} \\
\cmidrule(lr){3-5} \cmidrule(lr){6-8} \cmidrule(lr){9-11}
\textbf{Category} & \textbf{Outcome} & Coef. & 90\% CI & $q$ & Coef. & 90\% CI & $q$ & Coef. & 90\% CI & $q$ \\
\midrule
\endfirsthead

\toprule
\multicolumn{2}{c}{} & \multicolumn{3}{c}{\textbf{Visualization}} & \multicolumn{3}{c}{\textbf{Interview}} & \multicolumn{3}{c}{\textbf{Interaction}} \\
\cmidrule(lr){3-5} \cmidrule(lr){6-8} \cmidrule(lr){9-11}
\textbf{Category} & \textbf{Outcome} & Coef. & 90\% CI & $q$ & Coef. & 90\% CI & $q$ & Coef. & 90\% CI & $q$ \\
\midrule
\endhead

\bottomrule
\endfoot

\multirow{4}{*}{Commonalities} 
& I probably have things in common with the other participants whose experiences and opinions helped to create these proposals. & 0.60** & [0.16, 1.03] & 0.05 & 0.02 & [-0.43, 0.47] & 0.94 & -0.04 & [-0.58, 0.55] & 0.93 \\
& I probably share values, opinions, or experiences with others who were part of this process. & 1.13*** & [0.66, 1.61] & <.001 & 0.16 & [-0.33, 0.64] & 0.85 & -0.17 & [-0.68, 0.50] & 0.93 \\
& I probably have very little in common with other people who were part of this process. & 0.55 & [-0.10, 1.19] & 0.21 & -0.19 & [-0.85, 0.47] & 0.85 & -0.13 & [-1.01, 0.78] & 0.93 \\
& I am reading all of the questions in this survey carefully. & 0.02 & [-0.10, 0.14] & 0.78 & 0.04 & [-0.08, 0.17] & 0.85 & 0.01 & [-0.16, 0.18] & 0.93 \\

\multirow{4}{*}{General} 
& Think about yourself (Self) and the other participants in this study (Other)... & 0.89** & [0.22, 1.57] & 0.03 & 0.20 & [-0.49, 0.89] & 0.63 & -0.34 & [-1.30, 0.62] & 0.56 \\
& I feel connected to the other participants whose interviews were used to generate these proposals. & 1.39*** & [0.82, 1.97] & <.001 & 0.30 & [-0.29, 0.89] & 0.40 & -0.84* & [-1.66, -0.03] & 0.09 \\
& I would be willing to interact with the other participants whose interviews helped to generate the proposals. & -0.20 & [-0.68, 0.28] & 0.49 & 0.07 & [-0.42, 0.56] & 0.81 & 0.19 & [-0.48, 0.87] & 0.64 \\
& I don't feel a desire to learn more about the other participants whose interviews helped generate the proposals. & -0.18 & [-0.84, 0.48] & 0.65 & -0.58 & [-1.26, 0.09] & 0.16 & 0.11 & [-0.85, 1.04] & 0.85 \\

\multirow{4}{*}{Learn about others} 
& I learned more about others' experiences and opinions on this topic by interacting with the proposals. & 2.61*** & [2.08, 3.13] & <.001 & 0.94** & [0.40, 1.47] & 0.02 & -0.92** & [-1.65, -0.18] & 0.17 \\
& I understood other participants more holistically by interacting with the proposals. & 2.49*** & [1.94, 3.05] & <.001 & 0.62* & [0.05, 1.19] & 0.15 & -0.74 & [-1.53, 0.04] & 0.24 \\
& I was able to understand why other participants have certain beliefs or opinions on the topic. & 2.48*** & [1.91, 3.04] & <.001 & 0.45 & [-0.13, 1.02] & 0.27 & -0.42 & [-1.11, 0.38] & 0.39 \\
& This process did not clarify participants' experiences, beliefs, or opinions for me. & 3.07*** & [2.47, 3.68] & <.001 & 0.05 & [-0.57, 0.67] & 0.89 & -0.53 & [-1.39, 0.33] & 0.39 \\

\multirow{3}{*}{Perspective-taking} 
& I feel confident in predicting why different people might support or oppose a given proposal. & 0.33 & [-0.17, 0.84] & 0.28 & 0.05 & [-0.47, 0.57] & 0.87 & 0.23 & [-0.48, 0.95] & 0.86 \\
& I feel confident in identifying the key pros and cons of a given proposal. & 0.91*** & [0.45, 1.37] & 0.009 & 0.34 & [-0.14, 0.81] & 0.52 & -0.07 & [-0.72, 0.58] & 0.86 \\
& I would struggle to understand how someone with a different background or values might view each proposal. & 0.65* & [0.08, 1.23] & 0.09 & -0.34 & [-0.93, 0.25] & 0.52 & 0.22 & [-0.58, 1.04] & 0.86 \\

\multirow{3}{*}{Rational} 
& Are irrational for holding their stances on this topic & 0.84** & [0.21, 1.47] & 0.03 & -0.63 & [-1.28, 0.01] & 0.18 & 0.19 & [-0.71, 1.08] & 0.80 \\
& Have a stance that makes sense & 0.90*** & [0.40, 1.40] & 0.005 & -0.48 & [-0.99, 0.03] & 0.18 & -0.11 & [-0.81, 0.59] & 0.80 \\
& Are logical for holding their stance & 1.03*** & [0.52, 1.55] & 0.004 & -0.21 & [-0.74, 0.32] & 0.51 & -0.38 & [-1.11, 0.35] & 0.80 \\

\multirow{5}{*}{Respect} 
& I have respect for the other people whose experiences and opinions helped create these proposals. & 1.00*** & [0.58, 1.42] & <.001 & 0.03 & [-0.40, 0.46] & 0.90 & -0.60* & [-1.19, -0.01] & 0.29 \\
& Even if I disagreed with proposal decisions generated by this process, I still have respect for the people that support it. & 0.67** & [0.22, 1.12] & 0.02 & -0.12 & [-0.59, 0.34] & 0.82 & -0.41 & [-1.05, 0.23] & 0.29 \\
& Disregard the other participants' points of view & 1.13*** & [0.51, 1.75] & 0.008 & -0.18 & [-0.82, 0.45] & 0.82 & -0.71 & [-1.59, 0.17] & 0.29 \\
& Be considerate of the other participants' stances & 0.68*** & [0.29, 1.07] & 0.008 & 0.11 & [-0.29, 0.52] & 0.82 & -0.37 & [-0.92, 0.19] & 0.29 \\
& Take the other participants' points of view & 0.88** & [0.32, 1.44] & 0.01 & 0.34 & [-0.24, 0.92] & 0.82 & -0.54 & [-1.34, 0.25] & 0.29 \\

\multirow{5}{*}{Respect Pluralism} 
& I find it difficult to be open to political views that differ from my own. & 0.50 & [-0.07, 1.08] & 0.19 & -0.63* & [-1.22, -0.03] & 0.31 & 0.11 & [-0.70, 0.93] & 0.82 \\
& I respect people who disagree with me on political issues. & 0.48* & [0.08, 0.89] & 0.13 & 0.14 & [-0.28, 0.55] & 0.75 & -0.50 & [-1.07, 0.08] & 0.34 \\
& I would be willing to interact with someone who disagrees with me on political issues. & -0.02 & [-0.42, 0.38] & 0.93 & 0.13 & [-0.28, 0.54] & 0.75 & -0.36 & [-0.93, 0.21] & 0.37 \\
& It is important for people to respect a diversity of different opinions. & 0.62*** & [0.28, 0.96] & 0.02 & 0.33 & [-0.02, 0.68] & 0.31 & -0.42 & [-0.96, 0.07] & 0.34 \\
& If someone disagrees with me on a political issue, they probably have a good reason for doing so. & 0.45* & [0.02, 0.87] & 0.14 & 0.07 & [-0.37, 0.51] & 0.79 & -0.47 & [-1.08, 0.14] & 0.34 \\

\end{longtable}
}}
\subsection{Topic and Stances}
\begin{table}[H]
\centering
\footnotesize
\begin{tabular}{p{2cm}p{3cm}cccccccccc}
\toprule
\multicolumn{2}{c}{} & \multicolumn{3}{c}{\textbf{Visualization}} & \multicolumn{3}{c}{\textbf{Interview}} & \multicolumn{3}{c}{\textbf{Interaction}} & \\
\cmidrule(lr){3-5} \cmidrule(lr){6-8} \cmidrule(lr){9-11}
\textbf{Category} & \textbf{Outcome} & Coef. & 90\% CI & $p$ & Coef. & 90\% CI & $p$ & Coef. & 90\% CI & $p$ & $R^2_{adj}$ \\
\midrule
\multirow{4}{*}{Social Cohesion}
& Learning about topics 
    & 0.93*** & [0.63, 1.23] & 0.00 
    & 0.47 & [0.01, 0.16] & 0.78
    & -0.72** & [-1.15, -0.29] & 0.01
    & 0.28 \\
& Stance certainty 
    & -0.13 & [-0.33, 0.07] & 0.27
    & 0.10 & [-0.10, 0.39] & 0.30
    & 0.09 & [-0.19, 0.36] & 0.60
    & 0.33 \\
& Change stance (self report)
    & 0.32* & [0.02, 0.63] & 0.08
    & -0.17 & [-0.48, 0.39] & 0.15
    & -0.01 & [-0.44, 0.43] & 0.98
    & 0.12 \\
& Topic stance
    & 0.10 & [-0.07, 0.28] & 0.32
    & -0.04 & [-0.22, 0.68] & 0.13
    & -0.16 & [-0.40, 0.08] & 0.28
    & 0.50 \\
\bottomrule
\end{tabular}
\caption{Concept outcomes associated with attitude change across visualization and interview modes.}
\end{table}

{\footnotesize{
\begin{longtable}{p{1.8cm}p{3.5cm}ccccccccc}
\caption{Outcomes associated with attitude change across visualization and interview modes.} \\
\toprule
\multicolumn{2}{c}{} & \multicolumn{3}{c}{\textbf{Visualization}} & \multicolumn{3}{c}{\textbf{Interview}} & \multicolumn{3}{c}{\textbf{Interaction}} \\
\cmidrule(lr){3-5} \cmidrule(lr){6-8} \cmidrule(lr){9-11}
\textbf{Category} & \textbf{Outcome} & Coef. & 90\% CI & $q$ & Coef. & 90\% CI & $q$ & Coef. & 90\% CI & $q$ \\
\midrule
\endfirsthead

\toprule
\multicolumn{2}{c}{} & \multicolumn{3}{c}{\textbf{Visualization}} & \multicolumn{3}{c}{\textbf{Interview}} & \multicolumn{3}{c}{\textbf{Interaction}} \\
\cmidrule(lr){3-5} \cmidrule(lr){6-8} \cmidrule(lr){9-11}
\textbf{Category} & \textbf{Outcome} & Coef. & 90\% CI & $q$ & Coef. & 90\% CI & $q$ & Coef. & 90\% CI & $q$ \\
\midrule
\endhead

\bottomrule
\endfoot

Learning about topics & Compared to before, I feel more informed about why people may have different beliefs on raising the minimum wage. & 1.84*** & [1.19, 2.49] & 0.00 & 1.13** & [0.46, 1.79] & 0.03 & -1.47** & [-2.39, -0.55] & 0.02 \\
Learning about topics & I feel more informed about the topic of raising the minimum wage. & 1.71*** & [1.05, 2.38] & 0.00 & 0.80* & [0.12, 1.48] & 0.06 & -1.41** & [-2.36, -0.47] & 0.02 \\
Learning about topics & This exercise has taught me something about raising the minimum wage. & 1.62*** & [0.96, 2.27] & 0.00 & 0.80* & [0.13, 1.47] & 0.06 & -1.05 & [-1.97, -0.12] & 0.07 \\
Learning about topics & Compared to before, I feel more informed about why people may have different beliefs on hiring locals over international applicants. & 2.05*** & [1.42, 2.68] & 0.00 & 0.80* & [0.16, 1.45] & 0.06 & -1.23* & [-2.12, -0.34] & 0.03 \\
Learning about topics & I feel more informed about the topic of hiring locals over international applicants. & 1.85*** & [1.20, 2.50] & 0.00 & 1.02** & [0.35, 1.68] & 0.03 & -1.68** & [-2.60, -0.76] & 0.01 \\
Learning about topics & This exercise has taught me something about hiring locals. & 1.86*** & [1.20, 2.52] & 0.00 & 0.82* & [0.14, 1.49] & 0.06 & -1.64** & [-2.58, -0.71] & 0.01 \\
Learning about topics & Compared to before, I feel more informed about why people may have different beliefs about using race and gender in hiring. & 1.88*** & [1.25, 2.52] & 0.00 & 0.98** & [0.33, 1.63] & 0.03 & -1.34** & [-2.23, -0.44] & 0.02 \\
Learning about topics & I feel more informed about the topic of using race and gender in hiring decisions. & 1.97*** & [1.32, 2.62] & 0.00 & 1.13** & [0.46, 1.79] & 0.03 & -1.76** & [-2.68, -0.84] & 0.01 \\
Learning about topics & This exercise has taught me something about using race and gender in hiring. & 1.55*** & [0.87, 2.22] & 0.00 & 0.79* & [0.10, 1.48] & 0.06 & -1.02 & [-1.98, -0.07] & 0.08 \\

Stance certainty & I hold strong beliefs about raising the federal minimum wage. & 0.15 & [-0.28, 0.58] & 0.86 & 0.31 & [-0.13, 0.75] & 0.73 & -0.24 & [-0.85, 0.37] & 0.66 \\
Stance certainty & I feel certain my stance on the minimum wage is correct. & -0.04 & [-0.52, 0.44] & 0.89 & 0.00 & [-0.49, 0.50] & 0.99 & 0.28 & [-0.40, 0.96] & 0.66 \\
Stance certainty & I don't think there are good arguments against my minimum wage position. & -0.34 & [-0.96, 0.28] & 0.82 & -0.24 & [-0.88, 0.39] & 0.99 & 0.75 & [-0.13, 1.63] & 0.54 \\
Stance certainty & I hold strong beliefs about prioritizing locals over international applicants. & -0.36 & [-0.80, 0.08] & 0.77 & 0.08 & [-0.37, 0.53] & 0.99 & -0.32 & [-0.94, 0.30] & 0.66 \\
Stance certainty & I feel certain my stance on prioritizing locals is right. & -0.31 & [-0.75, 0.14] & 0.77 & -0.12 & [-0.58, 0.34] & 0.99 & 0.04 & [-0.59, 0.67] & 0.92 \\
Stance certainty & No good arguments exist against my position on prioritizing locals. & -0.72 & [-1.35, -0.10] & 0.53 & -0.09 & [-0.73, 0.56] & 0.99 & 0.72 & [-0.17, 1.61] & 0.54 \\
Stance certainty & Strong beliefs about using race/gender in hiring. & 0.07 & [-0.36, 0.51] & 0.89 & 0.63* & [0.18, 1.07] & 0.07 & -0.30 & [-0.92, 0.32] & 0.66 \\
Stance certainty & I feel certain about my stance on race/gender in hiring. & -0.07 & [-0.54, 0.40] & 0.89 & 0.52 & [0.04, 1.00] & 0.35 & -0.12 & [-0.78, 0.54] & 0.86 \\
Stance certainty & No good arguments exist against my stance on race/gender in hiring. & -0.22 & [-0.83, 0.38] & 0.86 & 0.05 & [-0.57, 0.67] & 0.99 & 0.82 & [-0.04, 1.68] & 0.54 \\

Topic stance & “The federal minimum wage should be raised to \$30 an hour.” & 0.01 & [-0.47, 0.49] & 0.97 & -0.23 & [-0.73, 0.27] & 0.80 & 0.18 & [-0.51, 0.86] & 0.67 \\
Topic stance & Prioritize hiring domestically before foreign applicants. & 0.39 & [-0.22, 1.00] & 0.85 & 0.10 & [-0.53, 0.72] & 0.80 & -0.79 & [-1.65, 0.08] & 0.40 \\
Topic stance & Race and gender should be used in hiring to combat inequality. & 0.20 & [-0.38, 0.79] & 0.85 & -0.18 & [-0.78, 0.42] & 0.80 & -0.28 & [-1.11, 0.55] & 0.67 \\

Change stance (self report) & My beliefs about the minimum wage have changed. & 0.19 & [-0.10, 0.49] & 0.28 & -0.07 & [-0.38, 0.23] & 0.68 & 0.04 & [-0.38, 0.45] & 0.98 \\
Change stance (self report) & My beliefs on hiring international applicants have changed. & 0.28* & [0.01, 0.55] & 0.18 & -0.10 & [-0.37, 0.18] & 0.68 & -0.05 & [-0.43, 0.33] & 0.98 \\
Change stance (self report) & My beliefs on using race/gender in hiring have changed. & 0.26 & [-0.02, 0.53] & 0.18 & -0.20 & [-0.48, 0.08] & 0.68 & 0.01 & [-0.38, 0.39] & 0.98 \\

\end{longtable}
}}

\section{AI Interviewer}
\label{app:ai-interviewer}

\subsection{AI Interview Questions}
This is the full set of initial questions that the AI interviewer asked participants along with the allocated times to answer. The interviewer continued to follow up until the maximum time was reached our three conversation turns were completed.
\label{app:int_q}
% Requires: \usepackage{longtable} and \usepackage{booktabs}
\setlength{\LTpre}{0pt}\setlength{\LTpost}{0pt}
\begin{center}
\footnotesize
\vspace{0.5em}
\begin{longtable}{p{7cm} p{7cm} r}
\toprule
\textbf{Question} & \textbf{Instruction} & \textbf{Max Sec} \\
\midrule
\endfirsthead
\toprule
\textbf{Question} & \textbf{Instruction} & \textbf{Max Sec} \\
\midrule
\endhead
\midrule
\multicolumn{3}{r}{\emph{Continued on next page}}\\
\endfoot
\bottomrule
\endlastfoot
To start off, can you tell me a bit about your background? Where did you grow up, and what was it like? You can share anything about your family, your neighborhood, friends, or anything else you can think of. & Learn as much as you can about the interviewee's life experience, and their personal background. Be respectful but curious as you hear about their story. Try to get some specific and detailed experiences or memories about the interviewee's life. & 30 \\
Now I'd like to ask you about an important relationship in your life. Who is someone who has had a big impact on you? Can you tell me about your relationship with them—how you met, what they mean to you, and any important memories you've shared? & Learn about a personally meaningful relationship. Encourage storytelling and emotional depth. Ask follow-ups to explore how this relationship shaped the interviewee's identity, values, or perspective. & 30 \\
Can you tell me about the work you do now? What's your job like, and how do you feel about it day to day? & Learn about what the interviewee does for a living. Be curious but respectful as you hear about their story. Follow up and ask what they like and dislike about their job. & 45 \\
Generally, what could be done to make your work life better? You can discuss whatever you feel would make your life better, either now or in the past. & Learn about what the interviewee thinks could be done to make their work life better. Be curious about their thoughts. Try to get some specific examples or ideas about how to improve their work life. If they need inspiration, you can give them some ideas but try not to bias them. If their answers are too brief or surface-level, probe further to understand their reasoning. You can even ask them what they would tell someone in-power like a lawmaker or employer. & 60 \\
We're now going to turn to the topic of the minimum wage. How would changes in the minimum wage (either raising or lowering it) affect you, your family, or anyone else you know personally? & Learn about how the interviewee thinks changes in the minimum wage would affect them, their family, or anyone else they know personally. Be curious about their thoughts. Make sure you ask follow up questions to get more specific on how they would be impacted. If they say they would be negatively impacted, ask why they think that. If they do not think they would be impacted, prompt them to think about different people they interact with that may be impacted. & 60 \\
What do you think is a fair minimum wage in your area? Why? & Understand the interviewee's thoughts on what a fair minimum wage is. Be curious and try to get some specific and detailed thoughts about the interviewee's thoughts on what a fair minimum wage is. For example, you can ask them to think about rent, food, and other basic necessities in their area. & 45 \\
If you earned more, what would you do with the extra money? & Learn about how the interviewee would spend extra money. You can also ask how it would make them feel to have the extra money, and what goals it would allow them to achieve. & 45 \\
Do you have any concerns about raising the minimum wage to \$30 an hour? & Surface how the interviewee thinks about economic trade-offs and competing narratives. If they cannot think of any downsides, you can provide some examples, such as higher prices, less jobs, etc. & 45 \\
I'd now like to ask you about discrimination. Have you ever experienced discrimination? If so, can you tell me about it? & Learn about the interviewee's experiences with discrimination. Be curious and try to get some specific and detailed experiences or memories about the interviewee's experience with discrimination. Follow up with questions about whether they have experienced discrimination in the workplace or hiring. If they say they have never experienced discrimination, ask them if anyone they know has experienced discrimination. If they say they have not experienced discrimination nor have anyone they know has experienced discrimination, then ask them whether they think discrimination is a problem in society and why. & 60 \\
How would you feel if you knew that race or gender was a factor in a hiring decision that affected you? & Understand the interviewee's thoughts on how they would feel if they knew that race or gender was a factor in hiring decisions. Be curious and try to get some specific and detailed thoughts about the interviewee's thoughts on how they would feel if they knew that race or gender was a factor in hiring decisions. & 45 \\
Do you think race or gender should be taken into account for hiring decisions to combat gender and racial inequality? Why or why not? & Understand the interviewee's thoughts on whether race or gender should be taken into account for hiring decisions. Be curious and try to get some specific and detailed thoughts about the interviewee's thoughts on whether race or gender should be taken into account for hiring decisions. & 45 \\
We're now going to move to a new topic. Have you ever worked with or known someone at work from another country? What was your experience like? & Understand whether the interviewee has worked with or known someone from another country. If they have, ask them specifics questions about their experience working with them and what it was like. If they have not, ask them if they have friends or family who are from another country, or they themselves are from another country. If they know someone from another country, ask them about that relationship and what it was like. & 45 \\
Have you or someone close to you ever been impacted by immigration policy, especially around work or hiring? & Understand whether the interviewee has been impacted by immigration policy. If they have, ask them specifics questions about their experience and what it was like. If they have not, ask them if they have friends or family who have been impacted by immigration policy. Try to get specific experiences or memories about the interviewee's experience with immigration policy if they have been impacted. If they have not been impacted at all nor have anyone they know has been impacted, then the objective has been met. & 45 \\
Do you think it matters where someone is from when hiring for a job, as long as they’re qualified? Why or why not? & Understand the interviewee's thoughts on whether it matters where someone is from when hiring for a job, as long as they’re qualified. Try to understand the benefits and drawbacks they see. Also follow up by asking if there are any situations where nationality or immigration status should be considered. & 45 \\
How do you feel about the idea that companies should prioritize hiring local applicants over foreign applicants? & Understand the interviewee's thoughts on the idea that companies should prioritize hiring local applicants over foreign applicants. Follow up by asking them the benefits and drawbacks of this idea. & 45 \\
\end{longtable}
\captionof{table}{Interview Questions with Instructions and Maximum Time}
\label{tab:questions_long}
\end{center}

\subsection{AI Interviewer Details}
We used an AI interviewer interface based on the system introduced by Park et al. (2024) for collecting participant experiences used in the audio clips. The implementation was optimized for responsiveness, emphasizing end-to-end voice interaction to create the impression of a live conversational interview. Visually, the interviewer was represented by a central 2D sprite avatar (the agent “Isabella”), while participants were shown as an avatar at the bottom of the screen moving toward a goal marker as the session progressed. Participants could optionally customize their avatar before beginning.
Participants responded to interview prompts by speaking naturally. The system monitored speech in real time and inferred response completion by detecting silences longer than four seconds. After each response, it automatically transcribed the participant’s speech and produced the next interviewer turn, which was delivered using text-to-speech. 

\section{Effects of Audio and Profile Exploration}
\label{app:audio_results}

\begin{figure}[H]
    \centering
    \includegraphics[width=0.5\linewidth]{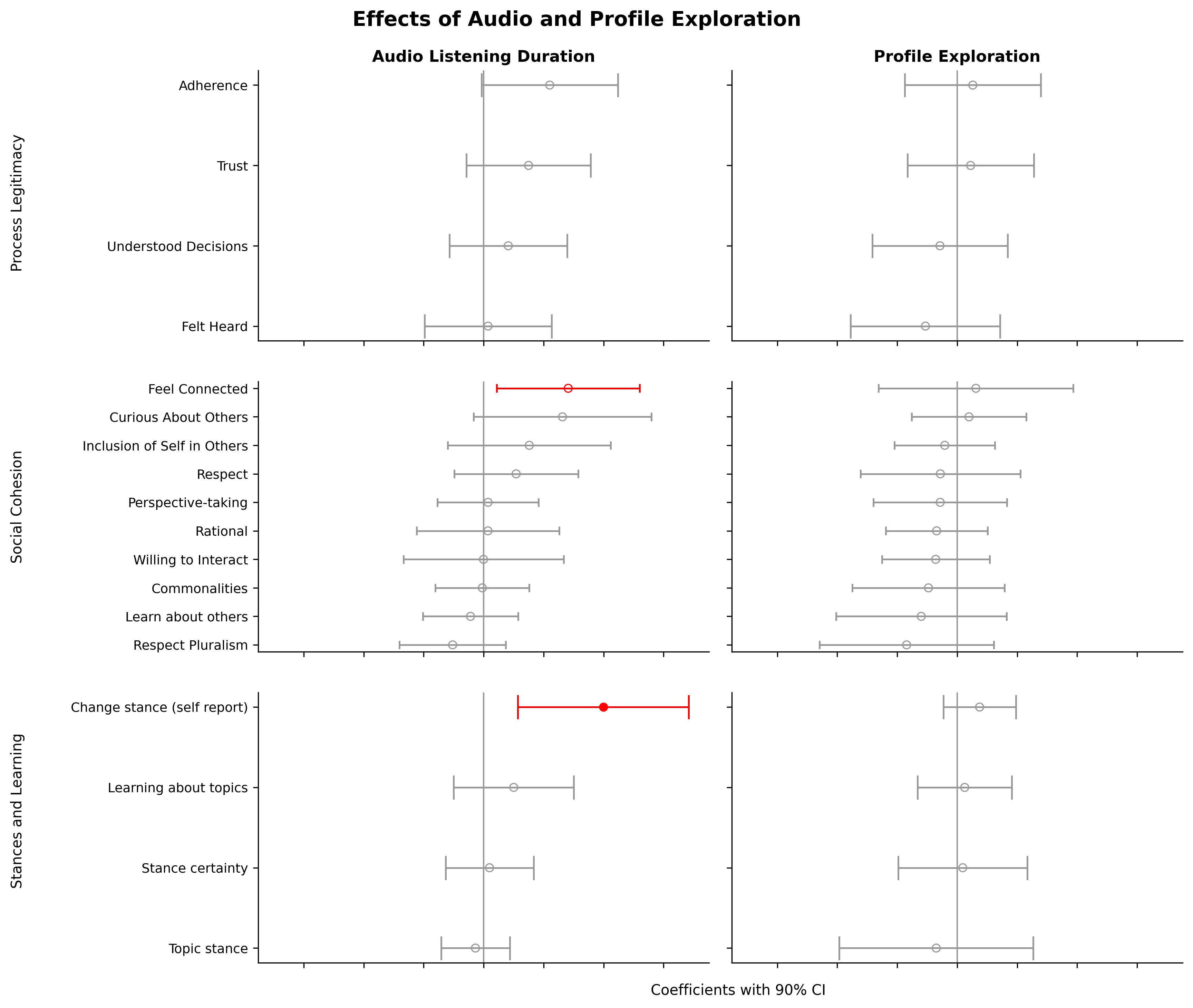}
    \caption{The red values indicate significance at \textit{p < 0.1}, with solid markers indicating significance at \textit{p < 0.05}.}
    \label{fig:audio_regression}
\end{figure}

\section{Participant Demographics}
\label{app:demographics}

\begin{table}[htbp]
\centering
\begin{tabular}{>{\raggedright\arraybackslash}p{0.16\linewidth}p{0.84\linewidth}}
\toprule
\textbf{Category} & \textbf{Distribution} \\
\midrule
Political Orientation & Very Liberal (12\%), Somewhat Liberal (8\%), Liberal (20\%), Moderate (16\%), Somewhat Conservative (9\%), Conservative (19\%), Very Conservative (15\%) \\
\addlinespace
Income & Less than \$30{,}000 (15\%), \$30{,}000--\$49{,}999 (15\%), \$50{,}000--\$99{,}999 (37\%), \$100{,}000 or more (31\%), Prefer not to answer (1\%) \\
\addlinespace
Education & Some college but no degree (15\%), Associate degree in college (2-year) (17\%), Bachelor's degree in college (4-year) (11\%), Master's degree (35\%), Doctoral degree (19\%), Professional degree (JD, MD) (1\%), Prefer not to answer (2\%) \\
\addlinespace
Ethnicity & Asian (7\%), Black (14\%), Mixed (6\%), White (69\%), Other (2\%), Unreported (3\%) \\
\addlinespace
Employment Status & Full-Time (50\%), Part-Time (19\%), Unemployed and seeking (8\%), Not in paid work (10\%), Due to start a new job (1\%), Other (3\%), Unreported (10\%) \\
\addlinespace
Sex & Female (59\%), Male (39\%), Unreported (2\%) \\
\addlinespace
Age & 18--30 (17\%), 31--45 (37\%), 46--59 (37\%), 60+ (9\%) \\
\bottomrule
\end{tabular}
\caption{We report the breakdown of our sample of participants across various social and demographic categories.}
\label{demographics}
\end{table}

\section{Interview Themes}
\label{app:themes}

\begin{table}[H]
\centering
\small
\begin{tabular}{>{\raggedright\arraybackslash}p{0.17\linewidth} p{0.33\linewidth} p{0.46\linewidth}}
\toprule
\textbf{Domain} & \textbf{Theme} & \textbf{Definition} \\
\midrule

\multirow{3}{*}{General Context}
  & Overall impressions of the process & Initial reactions, comfort, and general reflections on using the AI interviewer + platform \\
  & Tool role, usability, \& comprehension & Reflections on overall tool usability, its role, and features \\
  & Personal profile accuracy \& agency & Reactions to viewing their own profile \\
  & Profile representation & Views on how profiles were represented on the tool, including their positions on the 2-D plot \\
\cmidrule(l){2-3}

\multirow{9}{*}{Social Cohesion}
  &  {\textbf{Impact of “hearing” others}} & Impact of voices via transcripts and audio recordings \\
  &  {\textbf{Openness to engagement}} & Desire to interact further with other participants, and how \\
  &  {\textbf{Empathy building}} & Stated feelings of empathy regardless of stance (dis)agreement \\
  &  {\textbf{Shared values and commonalities}} & Similarities noticed with other profiles \\
  &  {\textbf{Barriers to connection}} & Reasons for lacking a sense of connection \\
  &  {\textbf{Impact of participants’ life experiences}} & Effects of seeing (or not seeing) lived experiences and personal stories \\
  &  {\textbf{Depth of participant explanations}} & Perceived depth behind others’ stated opinions \\
  &  {\textbf{Exposure to diverse perspectives}} & Seeing (or wanting to see) many viewpoints; evaluating issues from both sides \\
  &  {\textbf{Perceived rationality, understanding, and respect}} & Whether others’ views seemed rational/understandable/respectable \\
\cmidrule(l){2-3}

\multirow{4}{*}{Process Legitimacy}
  &  {\textbf{Participant voice in the process}} & Whether participants felt heard through this process \\
  &  {\textbf{Process clarity \& explanations}} & Feelings about black-box/opacity and features that would improve clarity and decision transparency (positive and negative) \\
  &  {\textbf{Conditional willingness to comply}} & Variability in accepting/following decisions; ties to legitimacy, norms, and consensus vs individual beliefs \\
  &  {\textbf{Future context for using this process}} & Contexts where participants would (not) use the tool (e.g., workplaces, elections, civic problems) \\
\cmidrule(l){2-3}

\multirow{3}{*}{Stances and Learning}
  &  {\textbf{Opinion shift}} & Accounts of stance change (or no change) \\
    & Perceptions of policy topics & Views on the choice of policy topics for this study \\
     & Nuance in policies & Mismatch between nuanced stances and binary outcomes \\
\cmidrule(l){2-3}

\multirow{4}{*}{Role of AI}
  &  {\textbf{AI use: opportunities}} & Positives from using AI in this process, including future potential \\
  &  {\textbf{AI vs human roles in process}} & Comparisons of human vs AI roles and autonomy \\
  &  {\textbf{AI use: concerns}} & Concerns (e.g., accuracy, reliability) \\
  &  {\textbf{Comfort and openness with the AI interviewer}} & Experiences with the AI interviewer \\

\bottomrule
\end{tabular}
\caption{Interview themes and their definitions mapped to our study's domains. The \textbf{bolded} themes are discussed in this paper for the purposes of the study.}
\label{tab:themes}
\end{table}

\section{Tool Screenshots}
\label{app:tool_screenshots}

\begin{figure}[httb]
    \centering
    \includegraphics[width=0.8\linewidth]{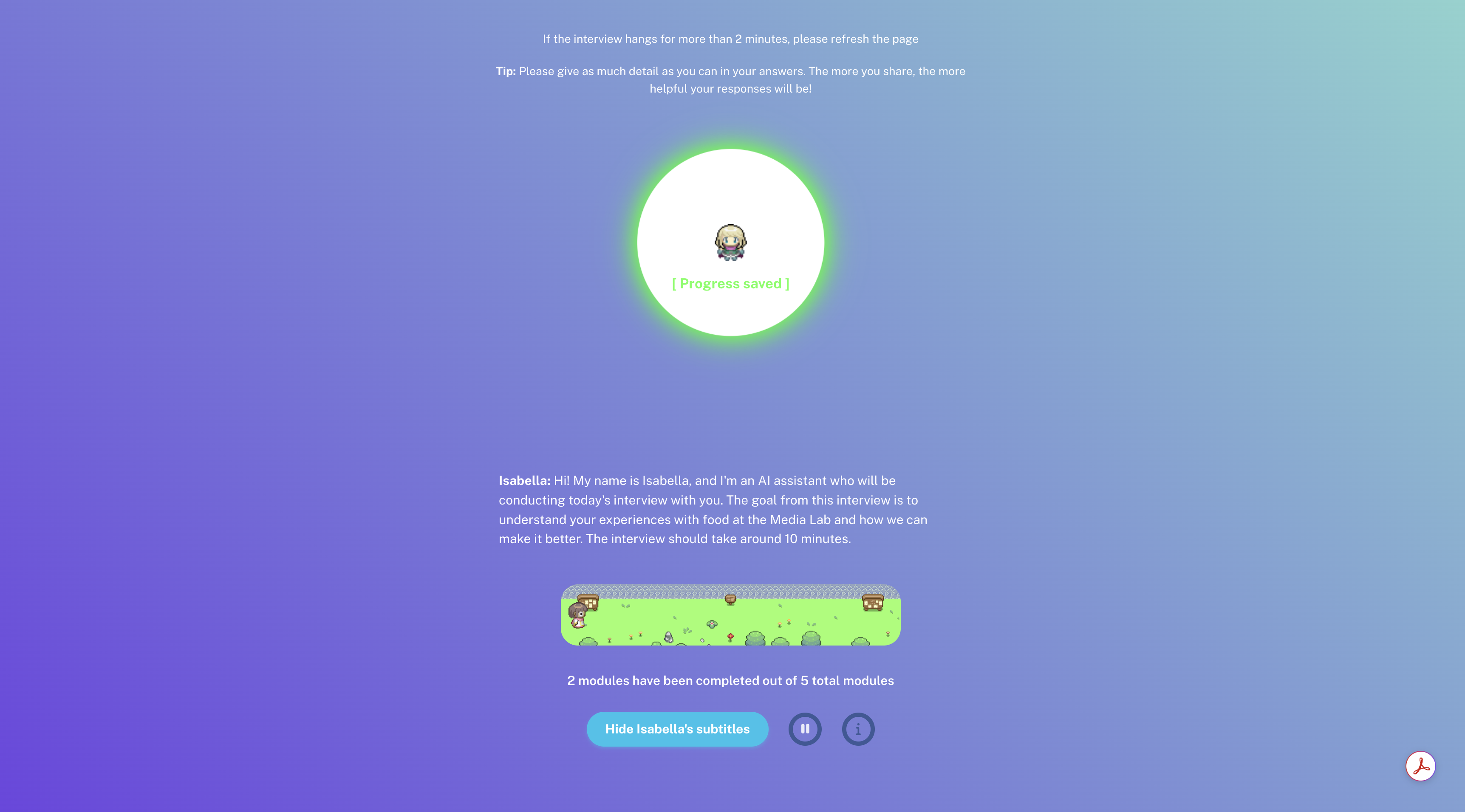}
    \caption{STEP 0: AI Interviewer screen where the user has a back n forth conversation with the AI agent on their personal life as well as policy related experiences}
    \label{fig:interviewer_screen}
\end{figure}

\begin{figure}[httb]
    \centering
    \includegraphics[width=0.9\linewidth]{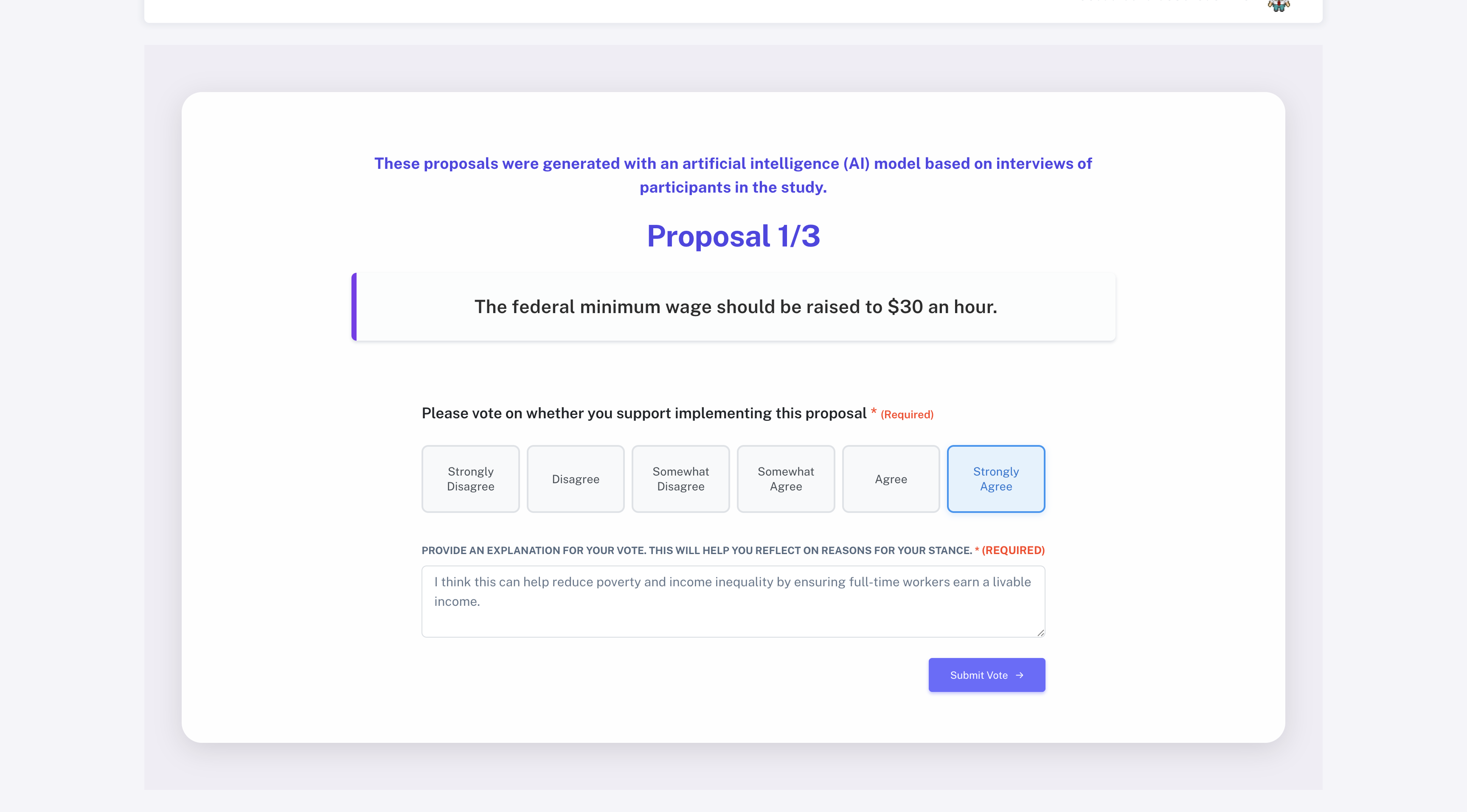}
    \caption{STEP 1: Voting screen where the user views the proposal and votes on it through a Likert scale and free-response question}
    \label{fig:voting_screen}
\end{figure}

\begin{figure}[httb]
    \centering
    \includegraphics[width=0.9\linewidth]{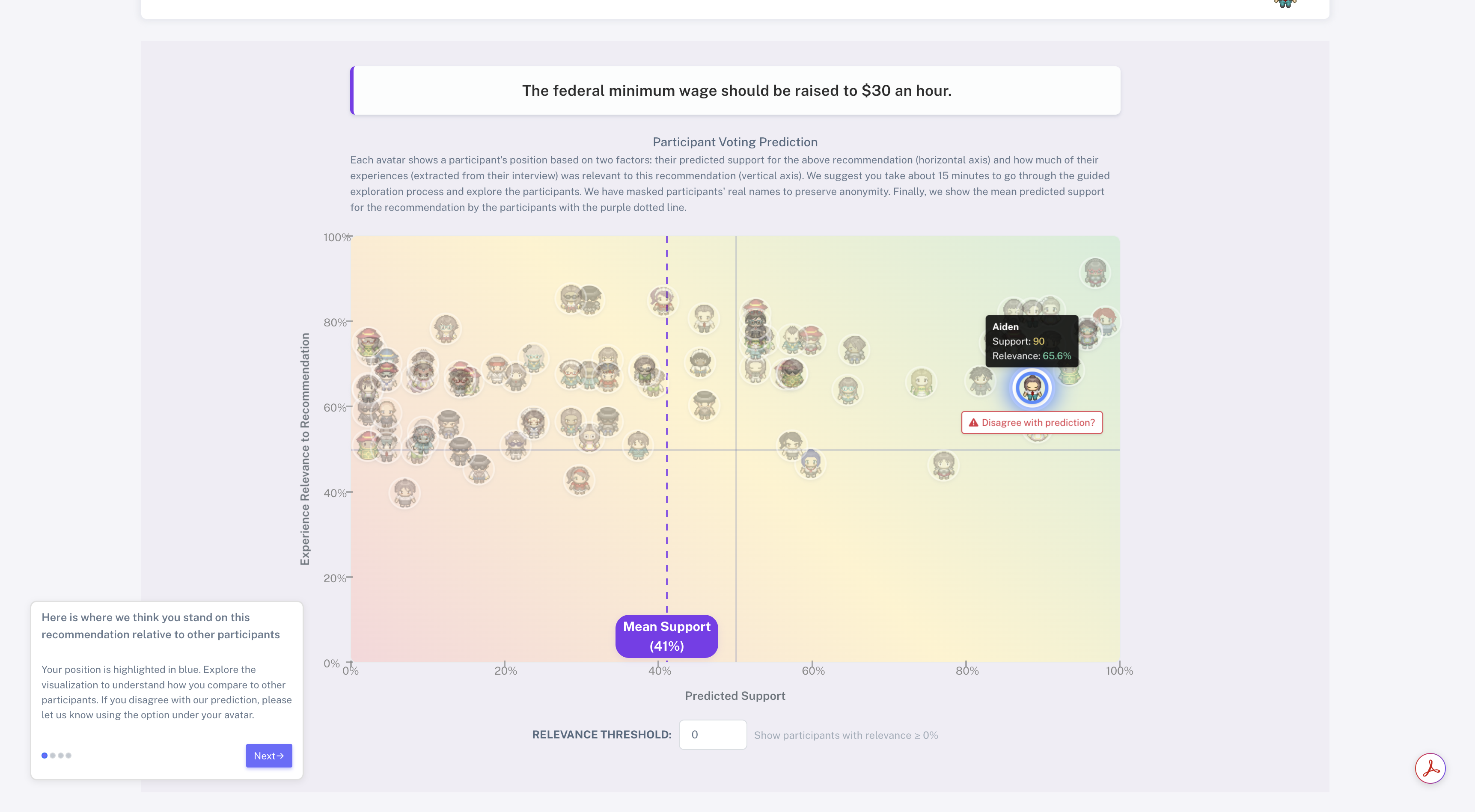}
    \caption{STEP 2a: Visualization showing the user where they stand relative to other study participants}
    \label{fig:placeholder}
\end{figure}

\begin{figure}
    \centering
    \includegraphics[width=0.9\linewidth]{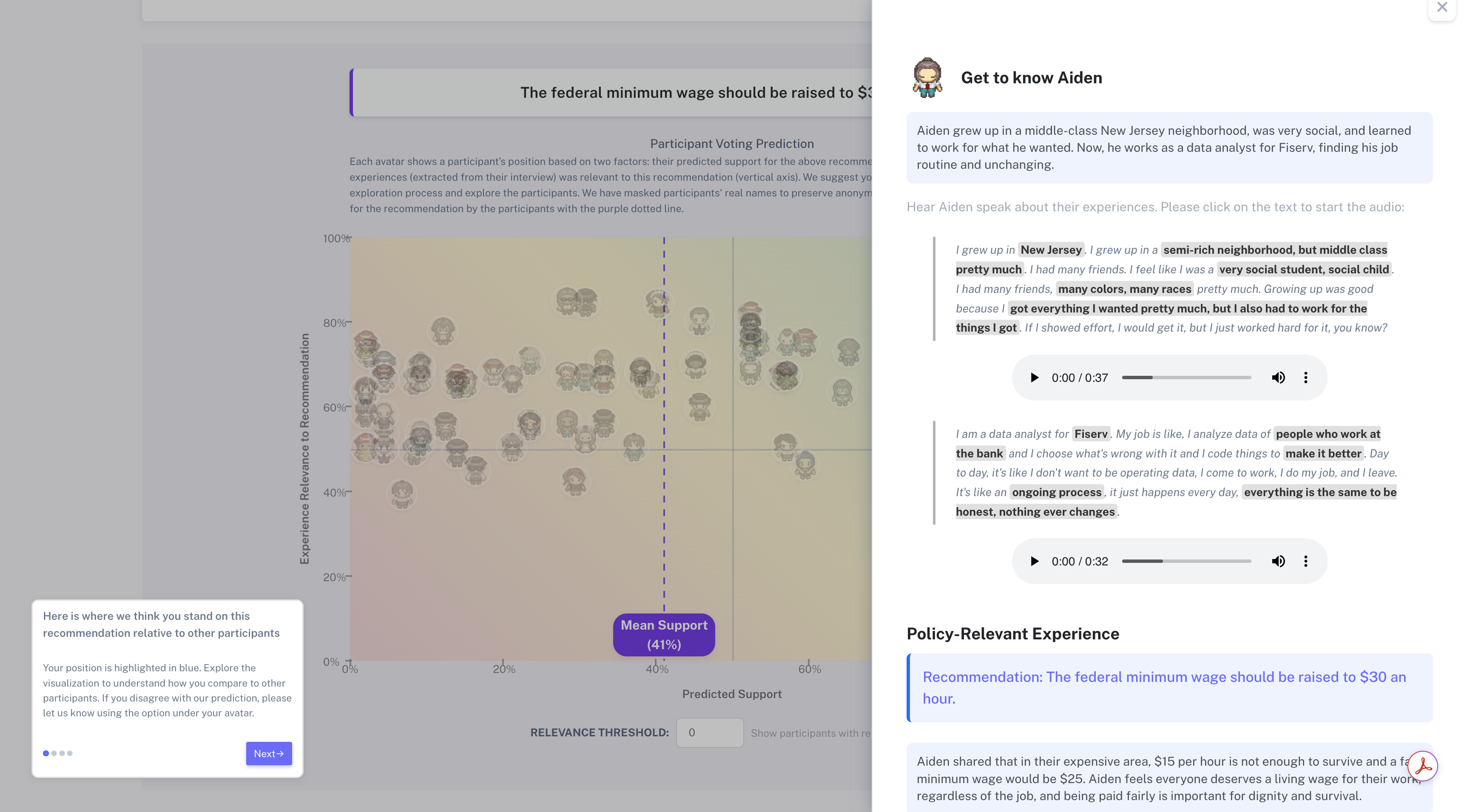}
    \caption{STEP 2b: Right side panel that opens up when the user clicks on their own avatar from the previous screen. This shows them how their life stories and experiences relevant to the proposal get extracted and shown to other participants}
    \label{fig:placeholder}
\end{figure}

\begin{figure}
    \centering
    \includegraphics[width=0.9\linewidth]{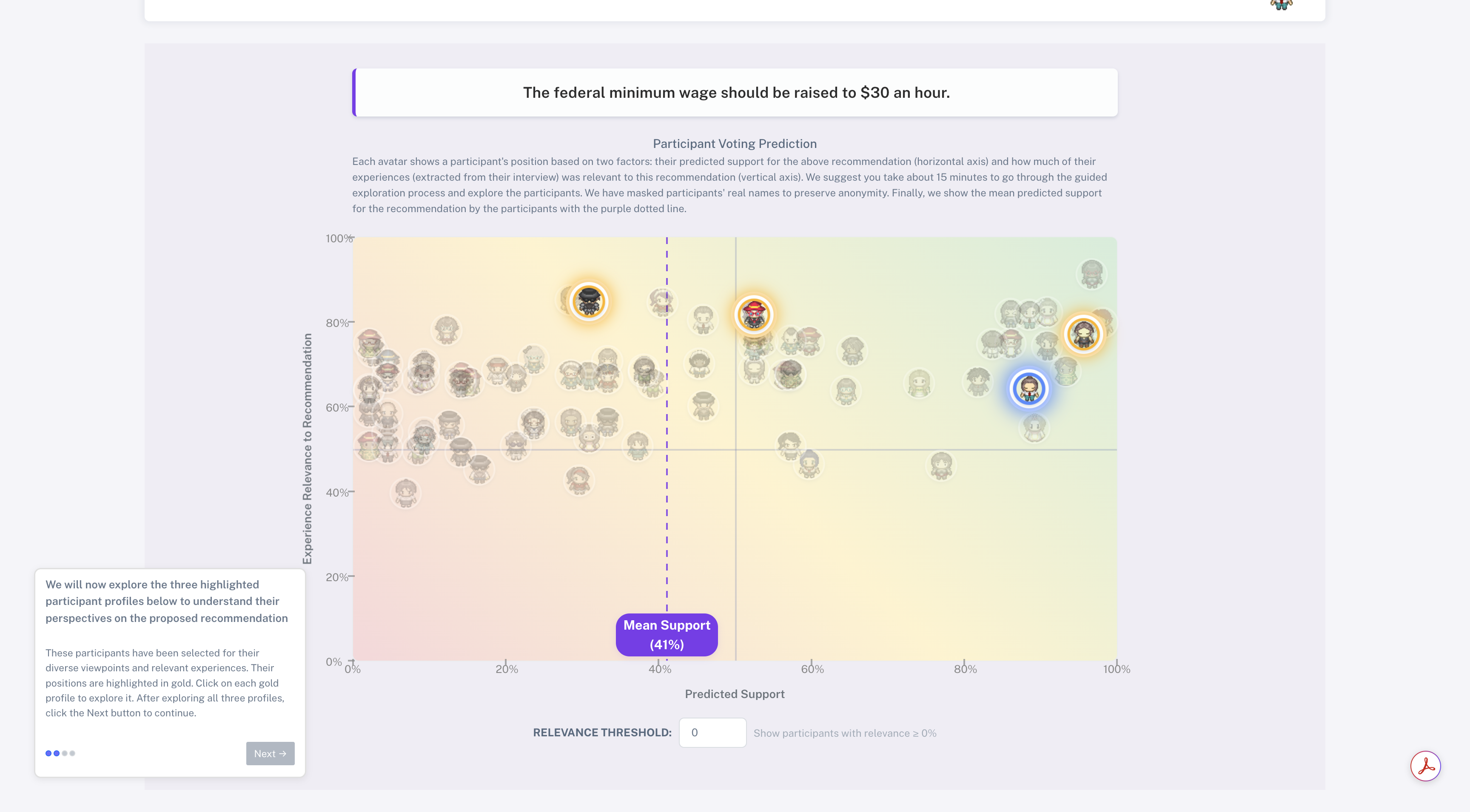}
    \caption{STEP 3a: Visualization showing three featured profiles of participants from across the spectrum on predicted support for the proposal}
    \label{fig:placeholder}
\end{figure}

\begin{figure}
    \centering
    \includegraphics[width=0.9\linewidth]{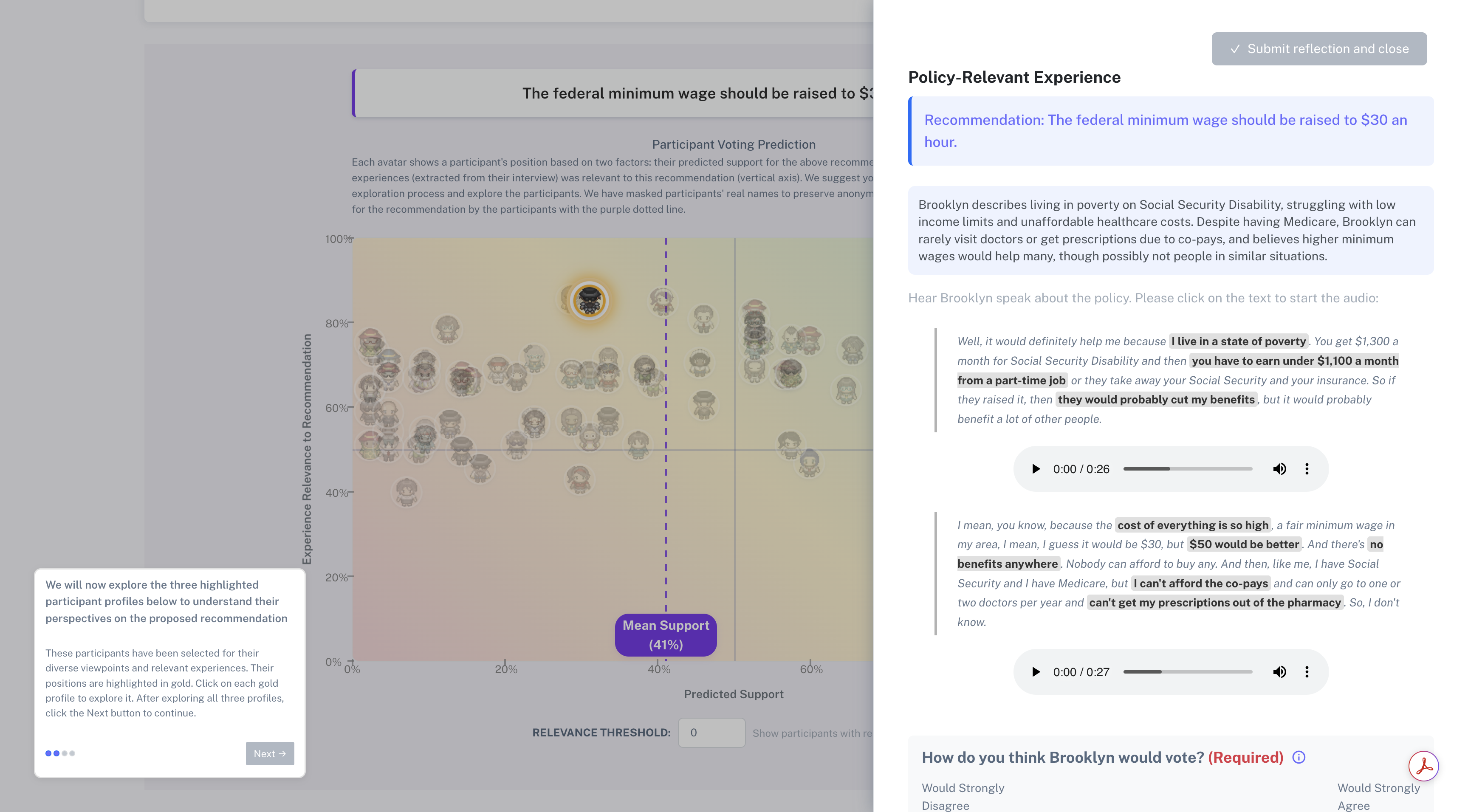}
    \caption{STEP 3b: Right side panel that opens up when the user clicks on any of the featured avatars from the previous screen. This allows the user to listen and react to the participant's life and policy-related experiences through with audio clips and reflection scaffolds}
    \label{fig:placeholder}
\end{figure}

\begin{figure}
    \centering
    \includegraphics[width=0.9\linewidth]{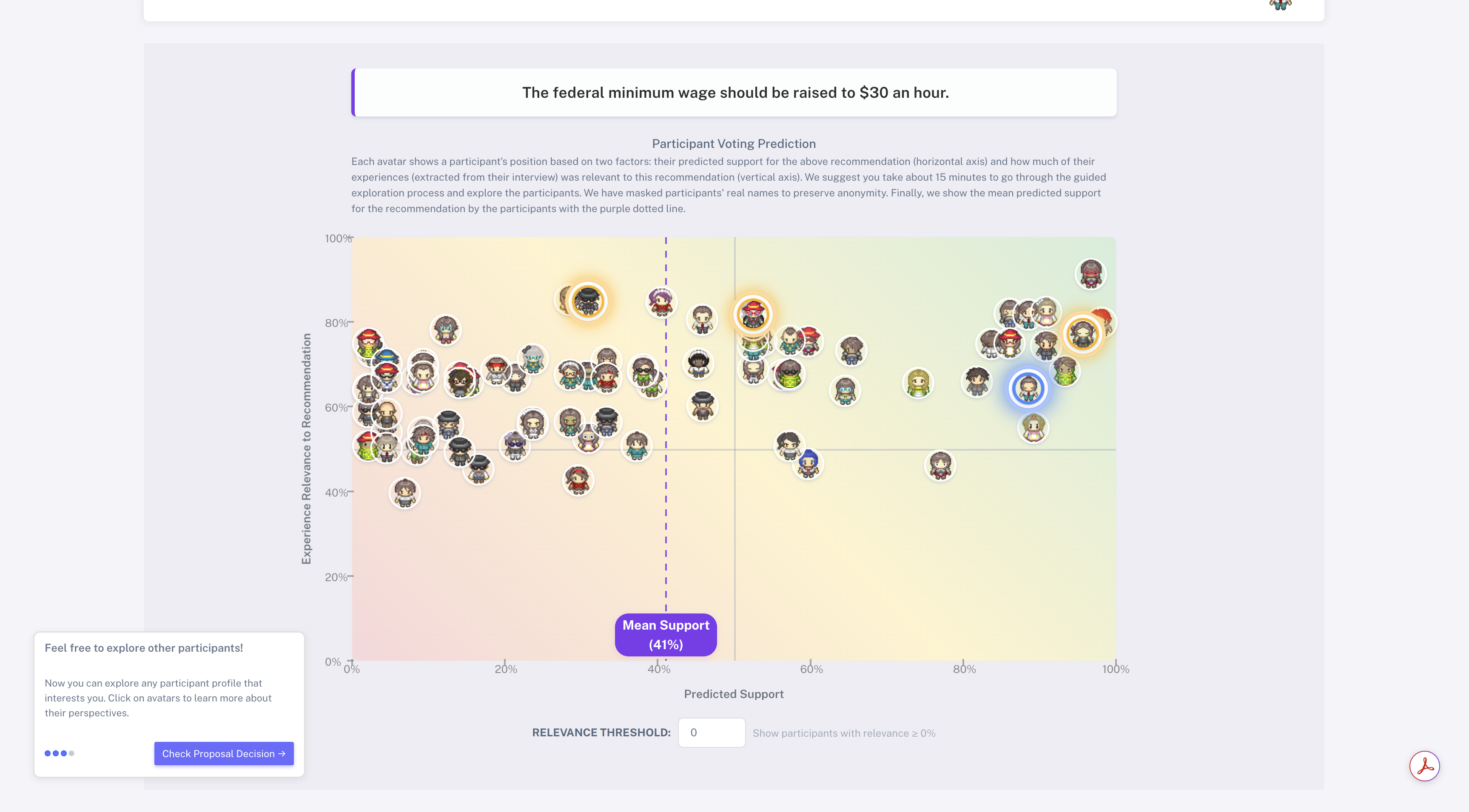}
    \caption{STEP 3c: Visualization showing all avatars on the spectrum. The user can optionally explore these after the previous steps}
    \label{fig:placeholder}
\end{figure}

\begin{figure}
    \centering
    \includegraphics[width=0.9\linewidth]{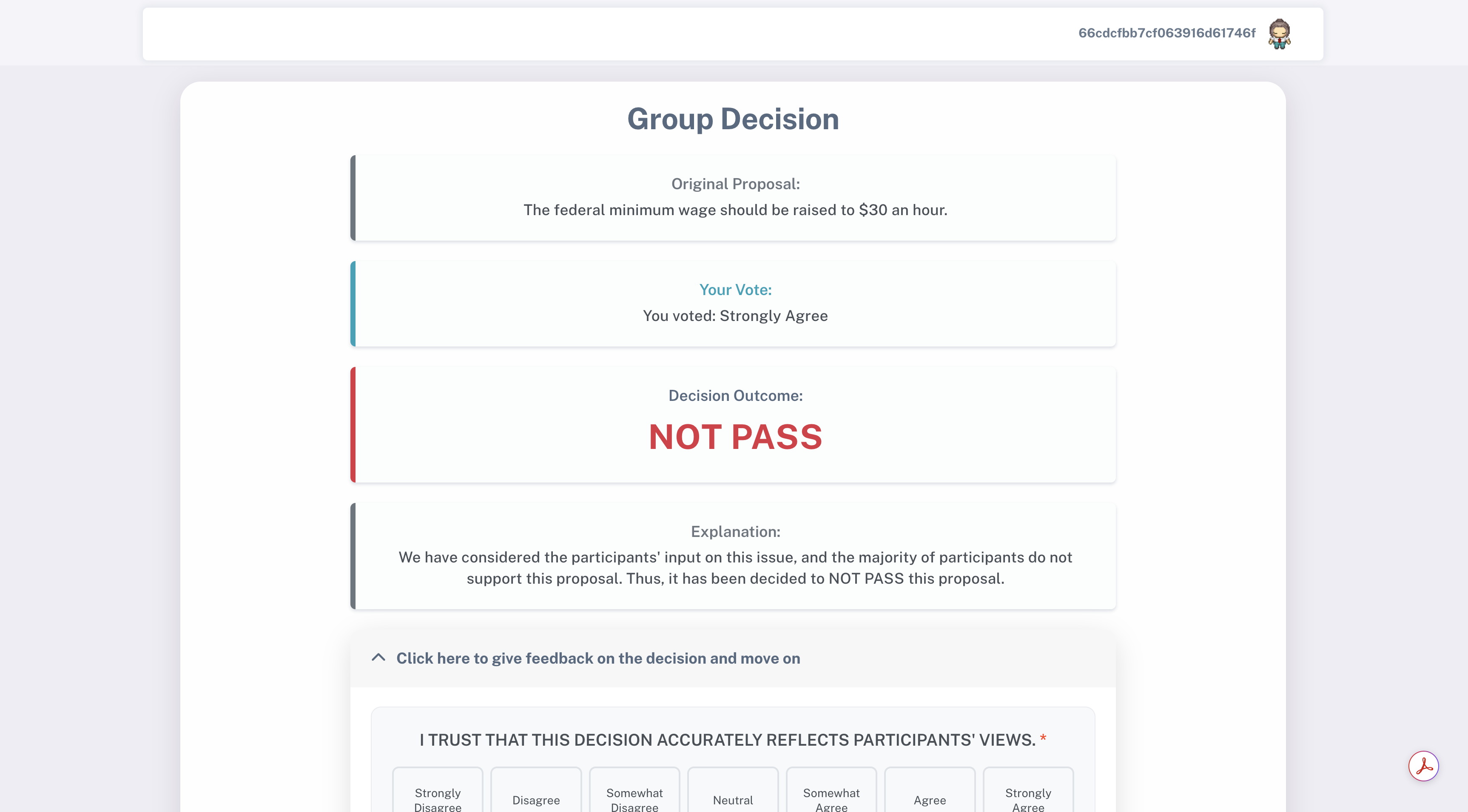}
    \caption{STEP 4: Decision page showing the user how the participant group voted summatively and if the proposal passed/failed. The user also gives feedback on the decision via likert scales below}
    \label{fig:placeholder}
\end{figure}

%TC:endignore

\end{document}